\documentclass[longauth]{aa} 
\usepackage{graphicx,natbib}
\usepackage{txfonts}
\usepackage{longtable}
\begin{document}
   \title{The long-lasting activity of \object{3C 454.3}}

   \subtitle{GASP-WEBT and satellite observations in 2008--2010\thanks{The radio-to-optical 
   data presented in this paper are stored in the GASP-WEBT archive; 
   for questions regarding their availability,
   please contact the WEBT President Massimo Villata ({\tt villata@oato.inaf.it}).}}

   \author{C.~M.~Raiteri              \inst{ 1}
   \and   M.~Villata                  \inst{ 1}
   \and   M.~F.~Aller                 \inst{ 2}
   \and   M.~A.~Gurwell               \inst{ 3}
   \and   O.~M.~Kurtanidze            \inst{ 4}
   \and   A.~L\"ahteenm\"aki          \inst{ 5}
   \and   V.~M.~Larionov              \inst{ 6,7,8}
   \and   P.~Romano                   \inst{ 9}
   \and   S.~Vercellone               \inst{ 9}
   \and   I.~Agudo                    \inst{10,11}
   \and   H.~D.~Aller                 \inst{ 2}
   \and   A.~A.~Arkharov              \inst{ 7}
   \and   U.~Bach                     \inst{12}
   \and   E.~Ben\'{i}tez              \inst{13}
   \and   A.~Berdyugin                \inst{14}
   \and   D.~A.~Blinov                \inst{ 6,8}
   \and   E.~V.~Borisova              \inst{ 6}
   \and   M.~B\"ottcher               \inst{15}
   \and   O.~J.~A.~Bravo~Calle        \inst{ 6}
   \and   C.~S.~Buemi                 \inst{16}
   \and   P.~Calcidese                \inst{17}
   \and   D.~Carosati                 \inst{18,19}
   \and   R.~Casas                    \inst{20,21}
   \and   W.-P.~Chen                  \inst{22}
   \and   N.~V.~Efimova               \inst{ 6,7}
   \and   J.~L.~G\'omez               \inst{10}
   \and   C.~Gusbar                   \inst{15}
   \and   K.~Hawkins                  \inst{15}
   \and   J.~Heidt                    \inst{23}
   \and   D.~Hiriart                  \inst{13}
   \and   H.~Y.~Hsiao                 \inst{24,22}
   \and   B.~Jordan                   \inst{25}
   \and   S.~G.~Jorstad               \inst{11,6}
   \and   M.~Joshi                    \inst{11}
   \and   G.~N.~Kimeridze             \inst{ 4}
   \and   E.~Koptelova                \inst{22,26}
   \and   T.~S.~Konstantinova         \inst{ 6}
   \and   E.~N.~Kopatskaya            \inst{ 6}
   \and   S.~O.~Kurtanidze            \inst{ 4}
   \and   E.~G.~Larionova             \inst{ 6}
   \and   L.~V.~Larionova             \inst{ 6}
   \and   P.~Leto                     \inst{16}
   \and   Y.~Li                       \inst{15}
   \and   R.~Ligustri                 \inst{27}
   \and   E.~Lindfors                 \inst{14}
   \and   M.~Lister                   \inst{28}
   \and   A.~P.~Marscher              \inst{11}
   \and   S.~N.~Molina                \inst{10}
   \and   D.~A.~Morozova              \inst{ 6}
   \and   E.~Nieppola                 \inst{ 5,29}
   \and   M.~G.~Nikolashvili          \inst{ 4}
   \and   K.~Nilsson                  \inst{29}
   \and   N.~Palma                    \inst{15,30}
   \and   M.~Pasanen                  \inst{14}
   \and   R.~Reinthal                 \inst{14}
   \and   V.~Roberts                  \inst{15}
   \and   J.~A.~Ros                   \inst{21}
   \and   P.~Roustazadeh              \inst{15}
   \and   A.~C.~Sadun                 \inst{31}
   \and   T.~Sakamoto                 \inst{32}
   \and   R.~D.~Schwartz              \inst{33}
   \and   L.~A.~Sigua                 \inst{ 4}
   \and   A.~Sillanp\"a\"a            \inst{14}
   \and   L.~O.~Takalo                \inst{14}
   \and   J.~Tammi                    \inst{ 5}
   \and   B.~Taylor                   \inst{11,34}
   \and   M.~Tornikoski               \inst{ 5}
   \and   C.~Trigilio                 \inst{16}
   \and   I.~S.~Troitsky              \inst{ 6}
   \and   G.~Umana                    \inst{16}
   \and   A.~Volvach                  \inst{35}
   \and   T.~A.~Yuldasheva            \inst{ 6}
 }
  
   \institute{
          INAF, Osservatorio Astronomico di Torino, Italy                                                     
   \and   Department of Astronomy, University of Michigan, MI, USA                                            
   \and   Harvard-Smithsonian Center for Astrophysics, Cambridge, MA, USA                                     
   \and   Abastumani Observatory, Mt. Kanobili, Abastumani, Georgia                                           
   \and   Aalto University Mets\"ahovi Radio Observatory, Kylm\"al\"a, Finland                                
   \and   Astron.\ Inst., St.-Petersburg State Univ., Russia                                                  
   \and   Pulkovo Observatory, St.-Petersburg, Russia                                                         
   \and   Isaac Newton Institute of Chile, St.-Petersburg Branch                                              
   \and   INAF-IASF Palermo, Italy                                                                            
   \and   Instituto de Astrof\'{i}sica de Andaluc\'{i}a, CSIC, Granada, Spain                                 
   \and   Institute for Astrophysical Research, Boston University, MA, USA                                    
   \and   Max-Planck-Institut f\"ur Radioastronomie, Bonn, Germany                                            
   \and   Instituto de Astronom\'{i}a, Universidad Nacional Aut\'{o}noma de M\'{e}xico, Mexico                
   \and   Tuorla Observatory, Univ.\ of Turku, Piikki\"o, Finland                                             
   \and   Astrophysical Inst., Department of Physics and Astronomy, Ohio Univ., Athens OH, USA                
   \and   INAF, Osservatorio Astrofisico di Catania, Italy                                                    
   \and   Osservatorio Astronomico della Regione Autonoma Valle d'Aosta, Italy                                
   \and   EPT Observatories, Tijarafe, La Palma, Spain                                                        
   \and   INAF, TNG Fundaci\'on Galileo Galilei, La Palma, Spain                                              
   \and   Institut de Ci\`encies de l'Espai (CSIC-IEEC), Spain                                                
   \and   Agrupaci\'o Astron\`omica de Sabadell, Spain                                                        
   \and   Graduate Institute of Astronomy, National Central University, Taiwan                                
   \and   ZAH, Landessternwarte Heidelberg-K\"onigstuhl, Germany                                              
   \and   Lulin Observatory, National Central University, Taiwan                                              
   \and   School of Cosmic Physics, Dublin Institute For Advanced Studies, Ireland                            
   \and   Department of Physics, National Taiwan University, Taipei, Taiwan                                   
   \and   Circolo Astrofili Talmassons, Italy                                                                 
   \and   Department of Physics, Purdue University, West Lafayette, IN, USA                                   
   \and   Finnish Centre for Astronomy with ESO (FINCA), University of Turku, Piikki\"o, Finland              
   \and   Facultad de Ciencias Espaciales, Univ.\ Nacional Autonoma de Honduras, Tegucigalpa M.D.C., Honduras 
   \and   Department of Physics, Univ.\ of Colorado Denver, CO, USA                                           
   \and   Center for Research and Exploration in Space Science and Technology, NASA/GSFC, Greenbelt, MD, USA  
   \and   Galaxy View Observatory, Sequim, Washington, USA                                                    
   \and   Lowell Observatory, Flagstaff, AZ, USA                                                              
   \and   Radio Astronomy Laboratory of Crimean Astrophysical Observatory, Ukraine                            
 }
   \date{}

  \abstract
   {The blazar \object{3C 454.3} is one of the most active sources from the radio to the $\gamma$-ray frequencies observed in the past few years.}
   {We present multiwavelength observations of this source from April 2008 to March 2010. The radio to optical data are mostly from the GASP-WEBT, UV and X-ray data from Swift, and $\gamma$-ray data from the AGILE and Fermi satellites. 
The aim is to understand the connection among emissions at different frequencies and to derive information on the emitting jet.}
   {Light curves in 18 bands were carefully assembled to study flux variability correlations. 
We improved the calibration of optical--UV data from the UVOT and OM instruments and estimated the Ly$\alpha$ flux to disentangle the contributions from different components in this spectral region.}
   {The observations reveal prominent variability above 8 GHz. 
In the optical--UV band, the variability amplitude decreases with increasing frequency due to a steadier radiation from both a broad line region and an accretion disc. 
The optical flux reaches nearly the same levels in the 2008--2009 and 2009--2010 observing seasons; the mm one shows similar behaviour, whereas the $\gamma$ and X-ray flux levels rise in the second period.
Two prominent $\gamma$-ray flares in mid 2008 and late 2009 show a double-peaked structure, with a variable $\gamma$/optical flux ratio. 
The X-ray flux variations seem to follow the $\gamma$-ray and optical ones by about 0.5 and 1 d, respectively. 
}
   {We interpret the multifrequency behaviour in terms of an inhomogeneous curved jet, where synchrotron radiation of increasing wavelength is produced in progressively outer and wider jet regions, which can change their orientation in time. In particular, we assume that the long-term variability is due to this geometrical effect. By combining the optical and mm light curves to fit the $\gamma$ and X-ray ones, we find that the $\gamma$ (X-ray) emission may be explained by inverse-Comptonisation of synchrotron optical (IR) photons by their parent relativistic electrons (SSC process). 
A slight, variable misalignment between the synchrotron and Comptonisation zones would explain the increased $\gamma$ and X-ray flux levels in 2009--2010, as well as the change in the $\gamma$/optical flux ratio during the outbursts peaks. The time delays of the X-ray flux changes after the $\gamma$, and optical ones are consistent with the proposed scenario.
}


   \maketitle
%

\section{Introduction}
A relativistic jet pointing at a small angle to the line of sight is most likely responsible for the extreme properties of the active galactic nuclei known as ``blazars", i.e.\ BL Lac objects and flat-spectrum radio quasars (FSRQ). Indeed, the alignment would cause Doppler enhancement of the emission and contraction of its variability time scales.
This peculiar orientation would also explain the apparent superluminal motion of radio knots in the jet. 
Owing to the jet emission beaming, we observe intensified synchrotron radiation from the radio to the UV--X-ray band and inverse-Compton radiation at higher energies, up to the TeV domain. 
Sometimes, in low brightness states, other contributions are detected, likely because of  unbeamed radiation from the blazar nucleus, i.e.\ the accretion disc and broad line region (BLR). This occurs more often in FSRQ than in BL Lac objects, and indeed the classical distinction between the two classes is based on the equivalent width of their broad emission lines (that has to be greater than 5 \AA\ in the rest frame for FSRQ).
One of the main issues concerns the origin of the seed photons that are Comptonised to X- and $\gamma$-ray energies: relativistic electrons certainly upscatter the soft photons they have previously produced by synchrotron emission (synchrotron-self-Compton, or SSC, process), but possibly also photons coming from outside the jet (external Compton, or EC, process), in particular from the accretion disc, the BLR, or an obscuring torus \citep[see e.g.][and references therein]{der09}. 

The quasar-type blazar \object{3C 454.3} has received particular attention by the international astronomical community since it underwent a major outburst in 2005 \citep{vil06,gio06,pia06,fuh06}. Indeed, after many decades of intense radio, but only mild optical activity, the source began brightening in the optical band in 2001, until in May 2005 it reached the maximum optical level ever observed, $R=12.0$. The outburst was simultaneously detected also in X-rays, while the millimetric radio flux peaked about one month after, and the 15--43 GHz flux much later \citep{vil07,rai08c}. According to \citet{vil06,vil07} the observed change in behaviour starting from 2001 was a geometric effect, i.e.\ the motion of a curved jet producing variations in the viewing angle of the different emitting regions.

In the following 2006--2007 observing season the source remained in a faint state, and new features appeared in its spectral energy distribution (SED). Through the analysis of low-energy data from the Whole Earth Blazar Telescope\footnote{\tt http://www.oato.inaf.it/blazars/webt/} (WEBT) and high-energy data from the XMM-Newton satellite, \citet{rai07b} were able to recognise both a little blue bump in the optical, possibly due to emission lines from the BLR, and a UV excess, suggesting a big blue bump likely due to thermal radiation from the accretion disc.
Moreover, they argue that the X-ray spectrum may be concave, with spectral softening at lower energies becoming more evident when the source is fainter, maybe revealing the high-frequency tail of the big blue bump.
Finally, they ascribed the flux excess in the $J$ band to a prominent broad H$\alpha$ emission line.
Both the UV excess and X-ray spectral curvature were also inferred from further XMM-Newton observations, at the beginning of the following observing season, while the contribution of the H$\alpha$ line was confirmed by near-IR spectroscopy at Campo Imperatore  \citep{rai08c}. 

Then, from July 2007 the source began a new activity phase, during which it was detected several times in $\gamma$-rays by the AGILE satellite\footnote{\tt http://agile.iasf-roma.inaf.it/}, and was monitored from the optical to the radio bands by the WEBT and its GLAST-AGILE Support Program \citep[GASP,][]{vil08}.
The results of these observations were published in
\citet{ver08}, \citet{rai08b,rai08c}, \citet{ver09}, \citet{don09b}, \citet{vil09b}, 
and \citet{ver10}.
Rotations of the optical polarisation vector both clockwise and counterclockwise were detected by \citet{sas10}.
In contrast, observations at very high energies ($\rm E > 100 \, GeV$) by the Cherenkov telescope MAGIC resulted in only upper limits \citep{and09}.

From the theoretical point of view, \citet{ghi07} compare the source SED of July 2007 with that of May 2005 and infer that the dissipation site in relativistic jets changes, with more compact emitting regions, smaller bulk Lorentz factors, and greater magnetic fields characterising locations closer to the black hole. However, in the \citet{ghi07} view, the dissipation always occurs in a subparsec zone, where the seed photons for Compton emission are provided by the BLR.
In contrast, \citet{sik08} claim that the dissipation region is located much farther, in the millimetric photosphere, at some parsecs from the central engine, so the Compton emission is due to scattering of infrared photons emitted by the hot dust.

In the meanwhile, the Fermi satellite\footnote{\tt http://fermi.gsfc.nasa.gov/} was launched, providing an unprecedented monitoring of 3C 454.3 at $\gamma$-ray energies \citep{bon09,abd09_3c454}.
The data from AGILE and Fermi were separately used to study the correlations between flux variations in optical and $\gamma$-rays. 
All results agreed that the time lag of $\gamma$-ray flux variations after the optical ones is less than one day \citep{ver09,don09b,bon09,ver10}.
\citet{fin10} discuss the spectral break at $\sim 2.4 \rm \, GeV$ shown by Fermi data in August 2008 suggesting that it results from the combination of Compton-scattered disc and BLR radiations.

An analysis of the radio--optical data in 2005--2007, including space observations by Spitzer in the infrared, revealed two synchrotron peaks, the primary at IR and the secondary at sub-mm wavelengths, likely arising from distinct regions in the jet \citep{ogl10}.

A multiwavelength study of the flaring behaviour of 3C 454.3 during 2005--2008 was performed by \citet{jor10}.
These authors suggest that the emergence of a superluminal knot from the core produces a series of optical and high-energy outbursts, and that the millimetre-wave core lies at the end of the jet acceleration and collimation zone. 

Suzaku observations of 3C 454.3 in November 2008 analysed by \citet{abd10_suzaku} confirmed the earlier suggestions by \citet{rai07b,rai08c} that there seems to be a soft X-ray excess, which  becomes more important when the source gets fainter. They interpret it as either a contribution of the high-energy tail of the synchrotron component or bulk-Compton radiation produced by Comptonisation of UV photons from the accretion disc by cold electrons in the innermost parts of the relativistic jet.

More recently, \citet[][see also \citealt{str10}]{pac10} have reported on a $\gamma$-ray flare of 3C 454.3 detected by AGILE in December 2009, when the peak flux reached $2.0 \times 10^{-5} \rm \, photons \, cm^{-2} \, s^{-1}$. The simultaneous flux increase at UV and optical frequencies was by far less dramatic. The authors found that the source behaviour during the flare cannot be accounted for by a simple one-zone model, but an additional particle component is needed. 
In contrast, \citet{bon11} explain the source behaviour in the same period by means of a one-zone model, where the magnetic field is slightly weaker when the overall jet luminosity is higher. 
A new intense $\gamma$-ray flare was detected by Fermi in April 2010 \citep[$\approx 1.6 \times 10^{-5} \rm \, photons \, cm^{-2} \, s^{-1}$,][]{ack10}, when fast variability on a time scale
of a few hours was revealed \citep{fos10}, supporting earlier results by \citet{tav10}, who claim that the dissipation region must be close to the black hole.
The December 2009 and April 2010 $\gamma$-ray outbursts detected by Fermi were further analysed by \citet{ack10}. These authors in particular discuss the observed spectral break at a few GeV, which they ascribe to a break in the underlying electron spectrum. This feature was also observed during the unprecedented $\gamma$-ray outburst of November 2010 \citep{abd11}, when the $\gamma$ flux rose to $(6.6 \pm 0.2) \times 10^{-5} \rm \, photons \, cm^{-2} \, s^{-1}$ showing clear spectral changes.

In this paper we present the radio-to-optical monitoring results obtained by the GASP-WEBT in 2008--2010, together with UV and X-ray data acquired by Swift and $\gamma$-ray observations by the AGILE and Fermi satellites.
We analyse the flux correlations at different frequencies to derive information on the jet physics and structure.
As in several previous WEBT-GASP papers, we focus on an interpretation of the observed light curves in terms of variations in the orientation and curvature of an inhomogeneous emitting jet.

\section{GASP-WEBT observations}
\label{sec_gasp}

   \begin{figure*}
   \centering
   \includegraphics{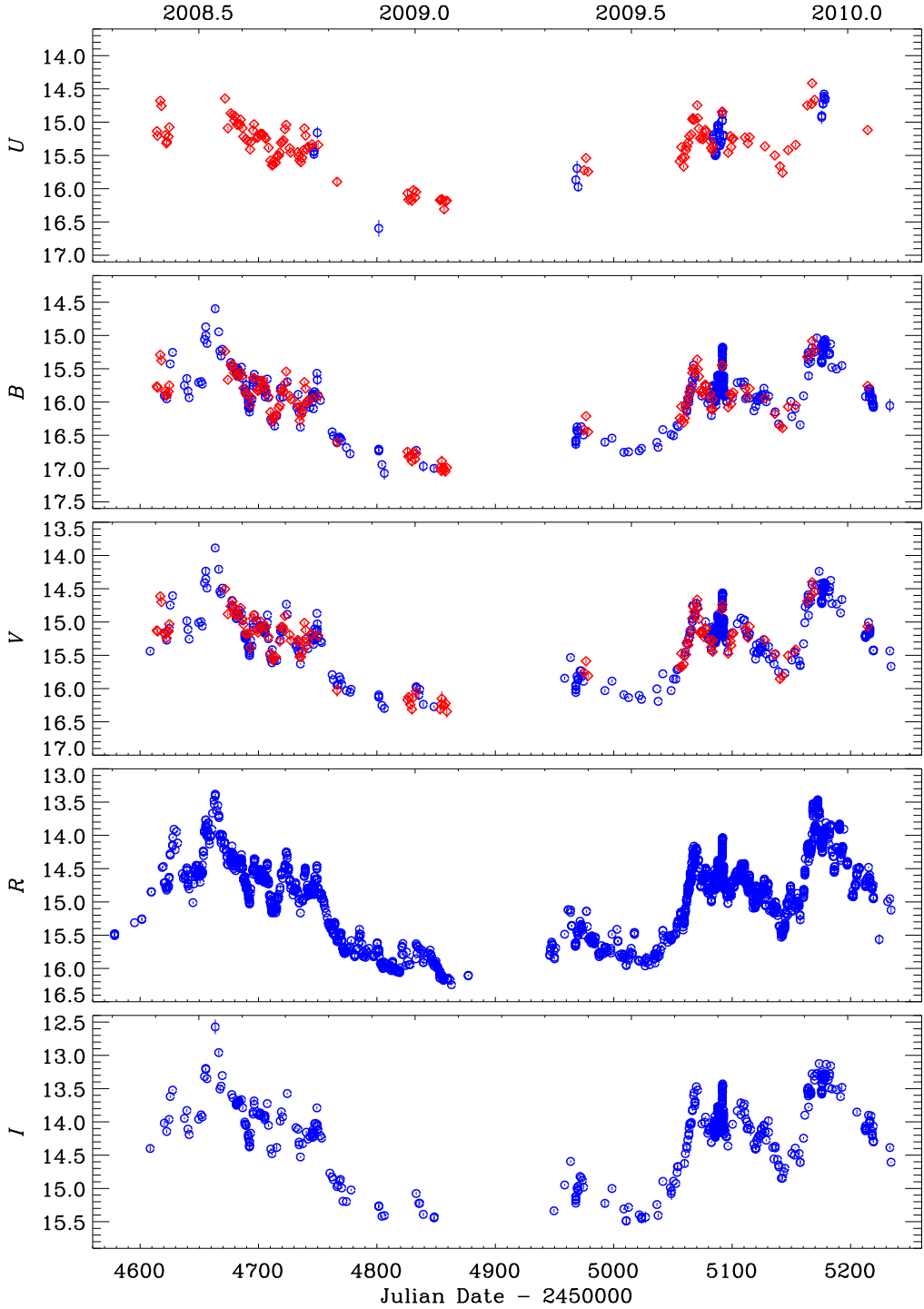}
   \caption{Optical light curves of 3C 454.3 in 2008--2010 built with GASP-WEBT data (blue circles) and Swift-UVOT data (red diamonds). The UVOT points have been shifted to match the ground-based data (see text for details).}
    \label{ottico}
    \end{figure*}

   \begin{figure*}
   \centering
   \includegraphics{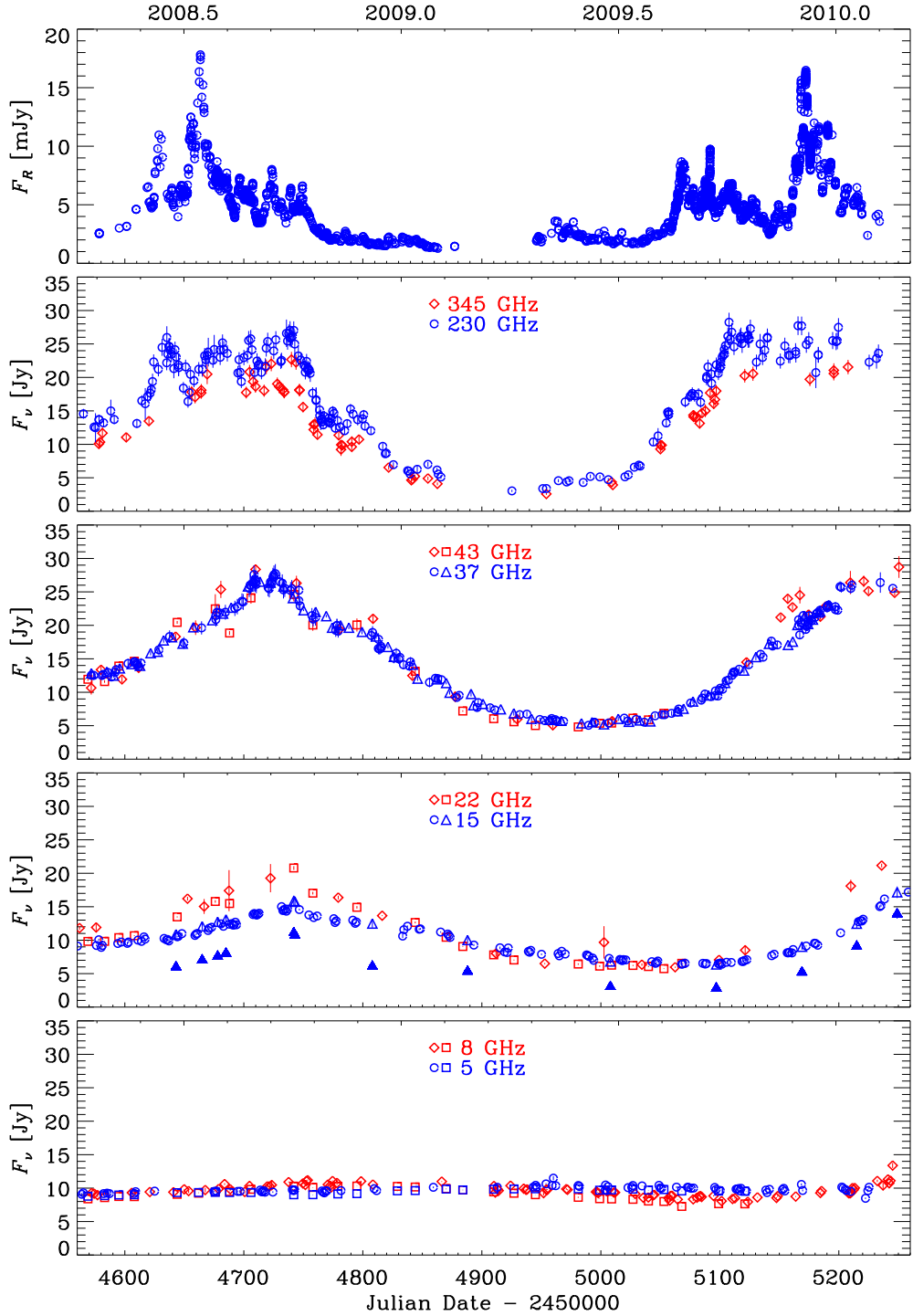}
   \caption{$R$-band optical flux densities in 2008--2010 (top) compared to the radio light curves at different frequencies. The GASP data (blue circles and red diamonds) are complemented with those from the Crimean Observatory at 37 GHz (blue triangles) and with those from the VLA/VLBA Polarization Calibration Database at 43, 22, 8 GHz (red squares), and at 5 GHz (blue squares). Data at 15 GHz from the MOJAVE Program are also included as blue triangles: the filled ones represent the core flux, and the empty ones the total flux.}
    \label{radop}
    \end{figure*}

The GLAST-AGILE Support Program (GASP) was born in 2007 as a project of the WEBT.
Its aim is to perform long-term multifrequency monitoring of selected $\gamma$-loud blazars during the operation of the AGILE and Fermi (formerly GLAST) $\gamma$-ray satellites.
In the optical, data are collected in the $R$ band, and photometric calibration of 3C 454.3 is obtained with respect to Stars 1, 2, 3, and 4 by \citet{rai98}. For this paper we also collected data taken by the GASP-WEBT observers in other optical bands to obtain spectral information.

Figure \ref{ottico} shows the optical light curves of 3C 454.3 in 2008--2010.
The 2008--2009 $R$-band light curve has already been published by \citet{vil09b}.
GASP-WEBT observations in 2009--2010 were performed at the following observatories:
Abastumani,
Calar Alto\footnote{Calar Alto data were acquired as part of the MAPCAT project: http://www.iaa.es/~iagudo/research/MAPCAT.},
Crimean,
Galaxy View,
Goddard (GRT),
Kitt Peak (MDM),
Lowell (Perkins),
Lulin,
New Mexico Skies,
Roque de los Muchachos (KVA),
Sabadell,
San Pedro Martir,
St.\ Petersburg,
Talmassons,
Teide (BRT),
Tijarafe, 
and Valle d'Aosta.
The figure also displays optical data acquired by the UVOT instrument onboard the Swift satellite (see Sect.\ \ref{sec_uvot}).

Both observing seasons are characterised by intense activity, with several flaring episodes superimposed on prominent outbursts. The total variation amplitude is similar in the two seasons and
it increases with increasing wavelength, as already noticed in previous works \citep[e.g.][]{vil06,rai08c}. This feature was interpreted as an effect of an additional, stabler, emission component, mainly contributing to the blue part of the optical spectrum. This radiation likely comes from the accretion disc and BLR (see also Sects.\ \ref{sec_uvot} and \ref{sec_niruv}).
Noticeable fast variability episodes are observed; in particular, a 0.3 mag brightening in about six hours on $\rm JD=2455066$ and 2455091, and a 0.5 mag brightening in about 14 hours, from $\rm JD=2455171.6$ to 2455172.2.

The radio light curves of 3C 454.3 at different wavelengths in 2008--2010 
are displayed in Fig.~\ref{radop}. 
Observations at 230 and 345 GHz are from the Submillimeter Array (SMA\footnote{These data were obtained as part of the normal monitoring programme initiated by the SMA \citep[see][]{gur07}}), a radio interferometer including eight dishes of 6 m size located atop Mauna Kea, in Hawaii.
Data at longer wavelengths were taken at the radio telescopes of Medicina (5, 8, and 22 GHz), Mets\"ahovi (37 GHz), Noto (43 GHz), and UMRAO (4.8, 8.0, and 14.5 GHz).
The Crimean Observatory provided additional observations at 37 GHz.
We also included data from the VLA/VLBA Polarization Calibration Database\footnote{{\tt http://www.vla.nrao.edu/astro/calib/polar/}. Data in the period between $\rm JD=2455110$ and 2455190 at 43 and 22 GHz, and between $\rm JD=2455130$ and 2455190 at 8 and 5 GHz were affected by problems in the automatic reduction procedure (Steven Myers, private communication) and were thus removed.} and from the MOJAVE programme \footnote{\tt http://www.physics.purdue.edu/astro/MOJAVE/}. 

In the top panel of Fig.\ \ref{radop} the $R$-band flux densities are shown for comparison. They were obtained by correcting magnitudes for a Galactic extinction of 0.29 mag and then converting them into fluxes using the absolute fluxes by \citet{bes98}.

At 230 GHz two outbursts with flickering superimposed cover the periods of high optical activity. Two outbursts are also visible at 43 and 37 GHz, but they are smoother and clearly delayed with respect to those at higher frequencies.

\section{Swift-UVOT observations}
\label{sec_uvot}
The Ultraviolet/Optical Telescope (UVOT; \citealt{rom05}) onboard the Swift spacecraft \citep{geh04} acquires data in the optical $v$, $b$, and $u$ bands, as well as in the UV filters $uvw1$, $uvm2$, and $uvw2$ \citep{poo08}. In the period from 2008 May 27 to 2010 January 17, Swift pointed 173 times at 3C 454.3, acquiring data with the UVOT instrument 168 times.
We reduced these data with the HEAsoft package version 6.10, with CALDB updated at the end of March 2010. 
Multiple exposures in the same filter at the same epoch were summed with {\tt uvotimsum}, and then aperture photometry was performed with the task {\tt uvotsource}. Source counts were extracted from a circular region with 5 arcsec radius, while background counts were estimated in two neighbouring source-free circular regions with 10 arcsec radii.
In this way we obtained 147, 143, 165, 128, 126, and 120 photometric data in the $uvw2$, $uvm2$, $uvw1$, $u$, $b$, and $v$ bands, respectively.

\begin{figure}
   \centering
   \resizebox{\hsize}{!}{\includegraphics{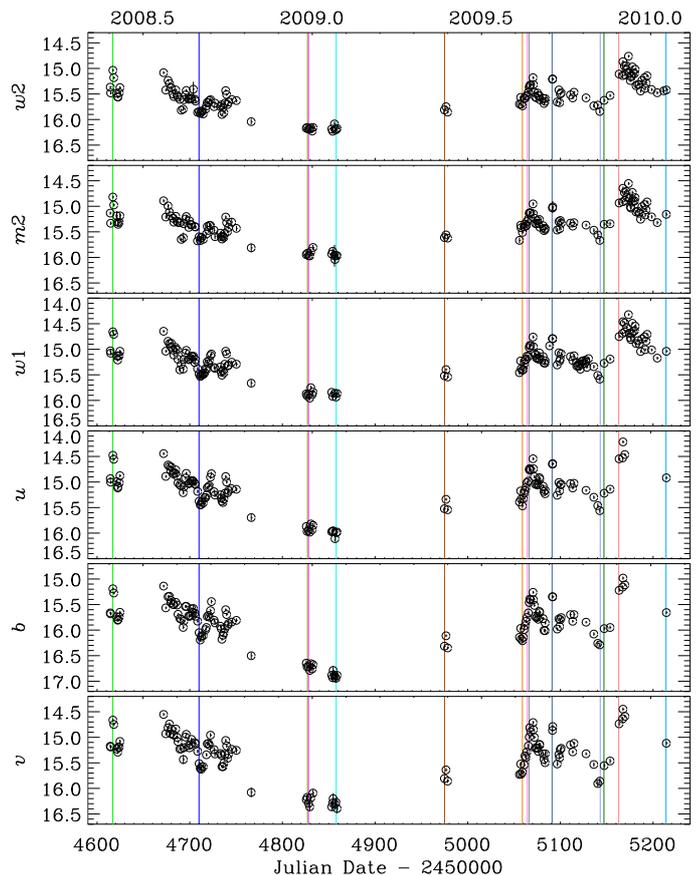}}
   \caption{Optical and UV light curves of 3C 454.3 in the 2008--2009 and 2009--2010 observing
seasons. All data were acquired with the UVOT instrument onboard the Swift satellite. Vertical lines indicate the epochs selected for the calibration procedure.}
    \label{uvot}
    \end{figure}

The UVOT light curves are shown in Fig.\ \ref{uvot}. 
The variability amplitude (maximum $-$ minimum magnitude) decreases with increasing frequency: 
1.94, 1.95, 1.88, 1.65, 1.48, and 1.46 mag going from the $v$ to the $uvw2$ band.
This confirms the trend already noticed for the optical data in Sect.\ \ref{sec_gasp} and extends it to the UV.

\begin{figure}
   \centering
   \resizebox{\hsize}{!}{\includegraphics{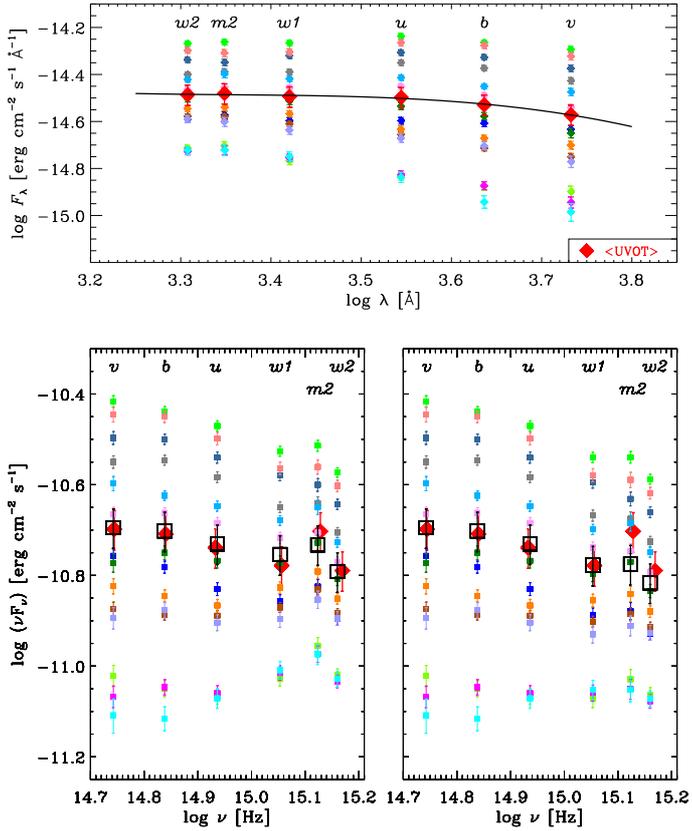}}
   \caption{Top: Observed spectra of 3C 454.3 in the optical--UV band obtained from different 
epochs of UVOT observations analysed in this paper. Red diamonds refer to the mean UVOT observed spectrum. The solid line is the log-cubic fit used in the calibration procedure.
Bottom-left: Optical--UV SEDs of 3C 454.3 corresponding to the observed spectra shown in the top panel. Red diamonds are derived from the UVOT average spectra shown in the top panel by using standard prescriptions to obtain de-reddened flux densities. Black squares represent the mean UVOT SED after recalibration as explained in the text.
Bottom-right: the same SEDs shown on the left, after correction for the emission lines contribution.}
    \label{calib}
    \end{figure}

The UVOT optical data are also reported in Fig.\ \ref{ottico}.
In this case, shifts have been given to the UVOT magnitudes to match the ground-based $U$, $B$, and $V$ data taken by the GASP-WEBT observers.
We estimated mean offsets $U-u=0.2$, $B-b=0.1$, and $V-v=-0.05$, with an uncertainty of a few hundredths of mag. These values are the same as those derived for BL Lacertae by \citet{rai10}.
By considering that the average UVOT colour indices of 3C 454.3 are: $u-b \sim -0.7$ and $b-v \sim 0.5$, the UVOT photometric calibrations by \citet{poo08} would indicate $U-u \sim 0.2$, $B-b \sim 0$--0.02, and $V-v \sim 0.01$--0.02. While our offset in the $u$ band agrees with their results, the offset we find in the other two bands is larger. 
This may partly depend on the uncertainties affecting the reference star photometry adopted for the calibration of the ground-based data.
Moreover, we must also consider the different spectral shape of our source with respect to the Pickles stars and GRB models used by \citet{poo08} to make the UVOT calibrations.
As a result, following \citet{rai10} we checked the calibrations for our source, calculating effective wavelengths, $\lambda_{\rm eff}$, and count rate to flux conversion factors, $\rm CF_\Lambda$, by folding them with both the blazar spectrum and the effective areas of the UVOT filters.
The same folding procedure was also applied to evaluate the amount of Galactic extinction for each band, $A_\Lambda$, starting from the \citet{car89} laws. This is an important point, because the average extinction curve shows a bump at about 2175 \AA, which makes the usual calculation of extinction at the filter $\lambda_{\rm eff}$ dangerous in the UV bands for very absorbed objects, as shown by \citet{rai10} for BL Lacertae.
Galactic reddening towards 3C 454.3 is smaller than in the BL Lacertae case, so we expect that deviations from both the \citet{poo08} calibrations and the Galactic extinction evaluated at the filter $\lambda_{\rm eff}$ are milder.
To verify this we selected a number of epochs where UVOT data were available in all filters and which represented different brightness levels of the source, and we built an average observed spectrum of 3C 454.3 (see Fig.\ \ref{calib}, top panel). Then we fitted this mean spectrum with a log-cubic curve and used this fit in the folding procedure.
We checked the stability of the results by iterating the procedure twice.

The new  $\lambda_{\rm eff}$, $\rm CF_\Lambda$, and $A_\Lambda$ are reported in Table \ref{caluvot}. For comparison, the corresponding ``standard" values ($\lambda_{\rm eff}$ and $\rm CF_\Lambda$ by \citealt{poo08}, and $A_\Lambda$ calculated at the $\lambda_{\rm eff}$ of \citealt{poo08}) are shown in brackets.
\begin{table}
\caption{Results of the UVOT recalibration procedure for 3C 454.3.}
\label{caluvot}      
\centering                          
\begin{tabular}{l c c c}       
\hline\hline                 
Filter & $\lambda_{\rm eff}$ & $\rm CF_\Lambda$ & $A_\Lambda$ \\    
       & \AA\ & $10^{-16} \rm \, erg \, cm^{-2} \, s^{-1} \, \AA^{-1}$ & mag \\
\hline                        
   $v$    & 5423 (5402) & 2.61 (2.614) & 0.36 (0.35)\\      
   $b$    & 4352 (4329) & 1.47 (1.472) & 0.47 (0.46)\\
   $u$    & 347{\bf 2} (3501) & 1.65 (1.63)  & 0.57 (0.55)\\
   $uvw1$ & 265{\bf 2} (2634) & 4.18 (4.00)  & 0.79 (0.73)\\
   $uvm2$ & 2256 (2231) & 8.42 (8.50)  & 0.99 (1.07)\\ 
   $uvw2$ & 2074 (2030) & 6.2{\bf 4} (6.2)   & 0.94 (1.02)\\ 
\hline                                  
\end{tabular}
\tablefoot{ Values in brackets represent ``standard" values, i.e.\ effective wavelengths $\lambda_{\rm eff}$ and count rate to flux conversion factors $\rm CF_\Lambda$ from \citet{poo08}, and Galactic extinction calculated from the \citet{car89} laws at the $\lambda_{\rm eff}$ by \citet{poo08}.
}
\end{table}
We then obtained recalibrated flux densities from count rates using the new $\rm CF_\Lambda$, and corrected for Galactic extinction according to the new $A_\Lambda$.
In the bottom-left panel of Fig.\ \ref{calib} we show the resulting SEDs corresponding to the selected observing epochs, together with their average. This average is compared to that obtained from the average spectrum shown in the top panel after applying standard calibrations and de-reddening.
The main differences are in the UV, where the recalibrated spectral shape is smoother.

As expected, there is a change in the spectral slope when going from bright to faint states, as
the ratio between the optical and UV fluxes decreases, leading to the already noticed smaller flux-density variation at higher frequencies. This reveals the presence of another emission component, likely coming from the accretion disc and BLR. Indeed,
BLR contributions to the source SED in the near-IR (H$\alpha$ line), in the optical band (\ion{Mg}{II}, \ion{Fe}{II}, and Balmer lines), and in the UV (Ly$\alpha$) have already been recognised \citep{rai07b,rai08c,benitez09,bon11}.
Subtraction of the Ly$\alpha$ flux would clarify to what extent a further contribution, possibly from the accretion disc, is needed.
We used the publicly available UV spectrum acquired by the Galaxy Evolution Explorer\footnote{\tt http://www.galex.caltech.edu/} (GALEX) on 2008 September 30 to estimate the Ly$\alpha$ flux, finding $\sim 1.8 \times 10^{-14} \rm \, erg \, cm^{-2} \, s^{-1}$, about 30\% lower than the value derived by \citet{wil95}. 
By folding the line profile with the effective areas of the UV filters, we found contributions of 
1.1, 3.1, and $1.2 \times 10^{-16} \rm \, erg \, cm^{-2} \, s^{-1} \, \AA^{-1}$ in the $uvw1$, $uvm2$, and $uvw2$ bands, respectively.
This means that in the $uvm2$ band the Ly$\alpha$ accounts for about 6\% of the source flux density in the brightest state shown in Fig.\ \ref{calib}, and for 17\% in the faintest state.
However, the Ly$\alpha$ line is not the only emission line to affect the UV spectrum of 3C 454.3.
The Hubble Space Telescope (HST) spectra acquired in 1991 and analysed by \citet{bah93}, \citet{wil95}, and \citet{eva04} also show prominent \ion{C}{IV} and \ion{O}{VI} + Ly$\beta$ lines, which are expected to mainly affect the flux in the $uvw1$ and $uvw2$ bands, respectively. 
We made a rough estimate of these further BLR contributions, finding that they are about one half of the Ly$\alpha$ ones.
The bottom-right panel of Fig.\ \ref{calib} shows the source SEDs already displayed in the bottom-left panel, after correction for the emission line flux contribution.
A UV excess still remains, likely indicating thermal emission from the accretion disc, as suggested by \citet{rai07b,rai08c} and \citet{bon11}.

\section{Spectral behaviour from the near-IR to the UV band}
\label{sec_niruv}

\begin{figure}
   \centering
   \resizebox{\hsize}{!}{\includegraphics{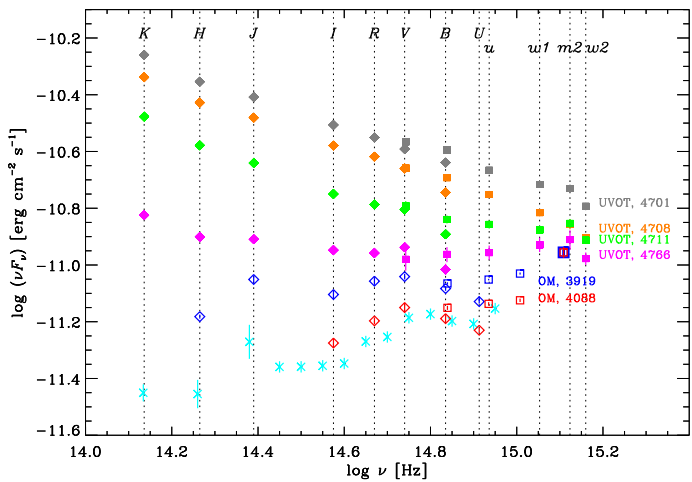}}
   \caption{SEDs of 3C 454.3 from the near-IR to the UV band at different brightness levels. The four SEDs plotted with filled symbols correspond to epochs of Swift observations: recalibrated UVOT data are displayed with squares, simultaneous optical and near-IR data from the GASP-WEBT with diamonds. The two SEDs shown with empty symbols correspond to the two XMM-Newton observations of July (blue) and December (red) 2006 \citep[see][]{rai07b}: recalibrated OM data are plotted with squares and WEBT data with diamonds. The epochs of satellite observations ($\rm JD - 2450000$) are indicated on the right. Error bars only take measure errors into account. Cyan crosses represent observations by \citet{neu79}.}
    \label{sed_iruv}
    \end{figure}

In the 2008--2009 observing season, near-IR monitoring in $J$, $H$, and $K$ bands was performed at Campo Imperatore. L'Aquila earthquake that occurred on 2009 April 6 prevented the acquisition of data in the following season. 
Figure \ref{sed_iruv} shows four near-IR-to-UV SEDs obtained with simultaneous near-IR and optical GASP-WEBT data and the Swift-UVOT data treated as explained in the previous section, but without subtraction of the emission line contributions. 
To add information on the source behaviour at even fainter levels, we reconsidered the two SEDs corresponding to the XMM-Newton observations of July and December 2006, which were published by \citet{rai07b}.
The corresponding near-IR and optical data were acquired by the WEBT collaboration, and they show the flux excess in the $J$ band owing to the H$\alpha$ emission line \citep[see also][]{rai08c} and the little bump in the optical, probably caused by \ion{Mg}{II}, \ion{Fe}{II}, and Balmer lines.
The OM data shown in Fig.\ 6 of \citet{rai07b} were treated in a standard way by converting magnitudes into fluxes according to the general method based on the Vega flux scale\footnote{\tt http://xmm.esa.int/sas/current/watchout/ Evergreen\_tips\_and\_tricks/uvflux.shtml} and by correcting for Galactic extinction calculated from the \citet{car89} law at the effective wavelength of the OM filters. We checked the reliability of those results by applying the same calibration procedure used in the previous section to check the UVOT calibrations.

Table \ref{calom} shows the results we obtained by folding the quantities of interest with both the source spectrum and the effective areas of the OM filters.
The most noticeable difference between the new and the standard $\lambda_{\rm eff}$ is a blueshift in the $B$ band, which makes the new value more like to the UVOT one. We also notice a 15\% increase in the $\rm CF_\Lambda$ in the $U$ band over Vega and a 10\% decrease in the Galactic extinction in the $UVW2$ band over the value we got from the \citet{car89} law at the standard effective wavelength.

\begin{table}
\caption{Results of the OM recalibration procedure for 3C 454.3.}
\label{calom}      
\centering                          
\begin{tabular}{l c c c}       
\hline\hline                 
Filter & $\lambda_{\rm eff}$ & $\rm CF_\Lambda$ & $A_\Lambda$ \\    
       & \AA\ & $10^{-16} \rm \, erg \, cm^{-2} \, s^{-1} \, \AA^{-1}$ & mag \\
\hline                        
   $V$    & 5423 (5430) & 2.53 (2.50) & 0.36 (0.35)\\      
   $B$    & 4340 (4500) & 1.35 (1.34) & 0.47 (0.46)\\
   $U$    & 3483 (3440) & 1.96 (1.70) & 0.56 (0.54)\\
   $UVW1$ & 2945 (2910) & 4.76 (4.86) & 0.67 (0.65)\\
   $UVM2$ & 2334 (2310) & 21.8 (21.9) & 0.96 (0.98)\\ 
   $UVW2$ & 2146 (2120) & 56.5 (58.8) & 1.00 (1.10)\\ 
\hline                                  
\end{tabular}
\tablefoot{Values in brackets represent, respectively, ``standard" \ effective wavelengths $\lambda_{\rm eff}$, count rate to flux conversion factors $\rm CF_\Lambda$ based on Vega, and Galactic extinction calculated from the \citet{car89} laws at the standard $\lambda_{\rm eff}$.
}
\end{table}

The main effect of the OM recalibration procedure is that the $U$ points in Fig.\ \ref{sed_iruv} are shifted to a higher flux level than in Fig.\ 6 of \citet{rai07b}, 
confirming what we have already found for the UVOT data, i.e.\ that the satellite data are much brighter than the ground-based one in the $U$ band. 
The recalibration of the OM data confirms the sharp rise of the SED in the UV when the source is in a faint state.

\section{Swift-XRT observations}
\label{sec_xrt}
In the period from 2008 May 27 to 2010 January 17 (see Sect.\ \ref{sec_uvot}) the XRT instrument onboard Swift \citep{bur05} performed 173 observations of 3C 454.3 in different observing modes.
We processed the event files acquired in pointing mode, using the HEASoft package version 6.10 with CALDB updated as 2010 September 30. 
We considered the observations with exposure time longer than five minutes, including 111 observations in photon-counting (PC) mode and 59 in windowed-timing (WT) mode. 
The task {\tt xrtpipeline} was run with standard filtering and screening criteria.
For the PC mode, we selected event grades 0--12. 
Source counts were extracted from a 30 pixel circular region ($\sim 71$ arcsec) centred on the source, and background counts were derived from a surrounding annular region with radii 110 and 160 pixels. When the count rate was greater than 0.5 counts/s, we punctured the source-extracting region, discarding the inner 3-pixel radius circle to correct for pile-up.
The {\tt xrtmkarf} task was used to generate ancillary response files (ARF), which account for different extraction regions, vignetting, and PSF corrections. The last in particular correct for the loss of counts caused by the central hole of the source extracting region in the pile-up case. 

For the data acquired in WT mode, we selected event grades 0--2. We used the same circular region of 30 pixel radius centred on the source to extract the source counts, and a similar region shifted away from the source along the window to extract the background counts.
When in the same observation the windows corresponding to different orbits overlapped with different position angles, we both created event files and extracted source and background counts separately for each orbit\footnote{We did not separate orbits if the misalignment was less than a few degrees.}. 
For each event file we then generated the exposure map with the task {\tt xrtexpomap}, which was used to obtain the ARF for the corresponding source spectrum.
Finally, we recombined the information from all the orbits of a single observation by summing the source spectra and the background spectra.  The ARFs were summed by weighting them according to their orbit contribution to the total counts of the observation.

The task {\tt grppha} was used to group the spectra and bin them to have more than 20 counts in each bin. These spectra were then analysed with the {\tt xspec} task, version 12.6.0q. We applied an absorbed single power-law model, where absorption is modelled according to \citet{wil00}.
Table \ref{xrt} presents the result of the spectral analysis in the case where the hydrogen column is fixed to $N_{\rm H}=1.34 \times 10^{21} \, \rm cm^{-2}$, as determined by the Chandra observations in 2005 \citep{vil06}.
Column 1 gives the ObsID,
Col.\ 2 the start time (UT), and
Col.\ 3 the total exposure time, often distributed over several orbits.
Column 4 gives the observing mode: an asterisk following ``PC" means that the count rate was greater than 0.5 counts/s, and the data were thus corrected for pile-up, while a number following ``WT" indicates how many misaligned windows were present in the total event file.
Column 5 gives the photon index $\Gamma$,
Col.\ 6 the 1 keV flux,
Col.\ 7 the value of the reduced $\chi^2$, while the number of degrees of freedom $\nu$ appears in brackets.

\addtocounter{table}{1}

Some observations resulted in a low number of counts, and the corresponding $\chi^2/\nu$ values were mostly small, indicating that the model is ``over-fitting" the data. 
Indeed, the $\chi^2$ statistic is not appropriate for a low number of counts, so that 
we refitted the spectra with either $\nu < 10$ or $\chi^2/\nu < 0.5$ adopting the Cash's C-statistic, which was developed to address this problem \citep{cas79,nou89}.
When in Col.\ 7 of Table \ref{xrt} the label ``Cash" appears, it means that the results of the spectral fit were obtained with the C-statistic.
The photon index $\Gamma$ ranges from 1.38 to 1.85, with an average value $< \Gamma > \, = 1.59$, and a standard deviation $\sigma = 0.09$. However, the mean square uncertainty $\delta^2=0.18$ is greater than the variance 
$\sigma^2$, so that the mean fractional variation \citep{pet01} $F_{\rm var}=\sqrt{\sigma^2-\delta^2}/ < \Gamma >$ is not even real. This means that the spectral variations are dominated by noise and cannot be ascribed to real changes in the source spectrum. The lack of a trend of $\Gamma$ with the source flux favours this interpretation.

\section{Flux correlations}
\label{flux_corr}

   \begin{figure*}
   \centering
   \includegraphics{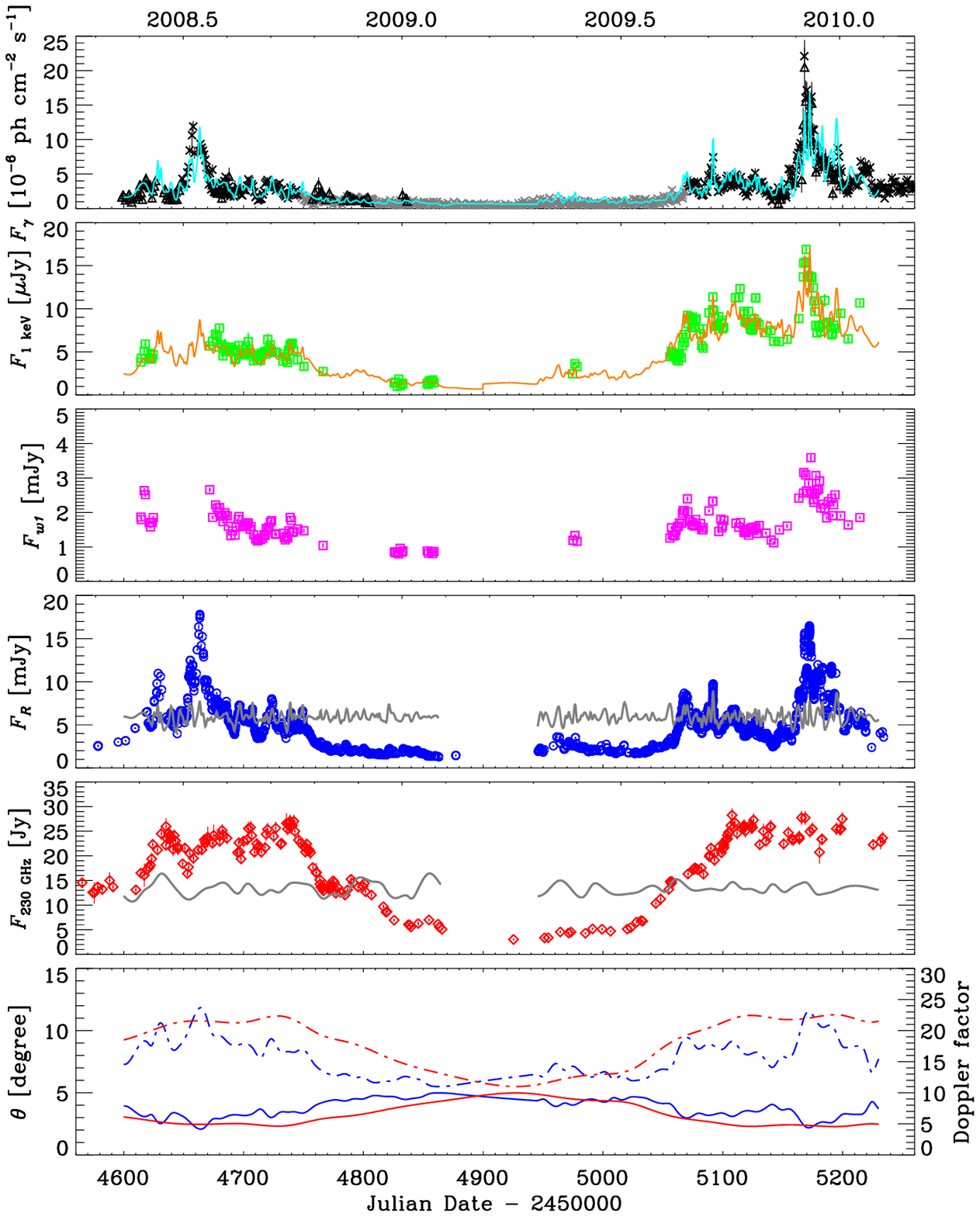}
   \caption{Light curves of 3C 454.3 from May 2008 to January 2010 at different frequencies. 
The $\gamma$-ray data are 0.1--300 GeV fluxes; black triangles are from AGILE \citep{ver10,pac10} and crosses from Fermi (the black ones from \citealt{abd09_3c454} and \citealt{ack10}, and the grey ones from the public light curve of the LAT monitored sources).
The 1 keV flux densities are derived from Swift-XRT spectra fitted with an absorbed power law with $N_{\rm H}=1.34 \times 10^{21} \, \rm cm^{-2}$ (see Sect.\ \ref{sec_xrt} and Table \ref{xrt}). De-reddened UV flux densities in the $w1$ band are obtained as explained in Sect.\ \ref{sec_uvot}. 
The optical $R$-band dereddened flux densities are derived from GASP data (see Sect.\ \ref{sec_gasp}). The mm light curve is from the SMA. 
The cyan and orange curves superposed on the $\gamma$- and X-ray light curves, respectively, are obtained by combining cubic spline interpolations through the 1-day binned optical and the 7-day binned mm light curves as explained in the text.
The bottom panel displays the evolution of the Doppler factors (dot-dashed lines) affecting the optical (blue) and mm (red) fluxes, and the viewing angles of the corresponding emitting regions (solid lines). Grey lines superposed on the optical and mm light curves represent intrinsic flux variations as would be observed under a constant Doppler factor $\delta = 18$.}
    \label{gamma}
    \end{figure*}

Figure \ref{gamma} shows the multiwavelength behaviour of 3C 454.3 from May 2008 to January 2010. 
The top panel displays the 0.1--300 GeV $\gamma$-ray light curve; the 2008--2009 AGILE data have already been published by \citet{ver10}, and the 2009--2010 ones by \citet{pac10}. Fermi-LAT data corresponding to the 2008 outburst are presented by \citet{abd09_3c454}, and those starting from August 2009 by \citet{ack10}. For the period in between we downloaded the public Fermi light curve from HEASARC as 3C 454.3 is part of the Fermi LAT Monitored Source List\footnote{\tt http://heasarc.gsfc.nasa.gov/W3Browse/fermi/\ fermilasp.html}. 
The 1 keV flux densities plotted in the second panel are the result of the Swift-XRT spectral fitting with an absorbed power law with fixed absorption discussed in Sect.\ \ref{sec_xrt} and reported in Table 3.
The following ultraviolet ($uvw1$ band), 
optical ($R$ band), and millimetric (230 GHz) light curves have already been shown in Figs.\ \ref{ottico} -- \ref{uvot}.
We notice a rough correlation among the fluxes at all wavelengths, but the variability amplitude decreases from the higher to the lower frequencies, with the ultraviolet the only exception.
Indeed, the maximum variation in the $\gamma$-ray flux is a factor $\sim 214$, while it is $\sim 17$ at 1 keV, $\sim 14$ in the optical $R$ band, and $\sim 9$ at $\sim 1 \, \rm mm$.
The total change of the $uvw1$ flux density is only a factor $\sim 5$, and this is the consequence of the dilution of the variable synchrotron emission with the less variable disc and BLR radiation (see Sects.\ \ref{sec_uvot} and \ref{sec_niruv}).

It is interesting to notice that the flux level reached by the optical light curve is similar in the two seasons; the same is true for the mm light curve, while in the X-ray and $\gamma$-ray bands there is a clear general flux increase in the second period. We miss information in the infrared, but can guess that the source behaviour would be intermediate between those in the optical and mm bands.

One possible interpretation of the observed long-term flux variations of 3C 454.3 at low frequencies (radio to UV) assumes that the jet is both inhomogeneous, with radiation of increasing wavelength being produced in progressively outer regions \citep[e.g.][]{ghi89,mar92}, and curved and that the alignment of the various emitting regions with the line of sight changes in time \citep{vil06,vil07,vil09b}.
Because the Doppler factor $\delta=[\Gamma (1- \beta \cos \theta)]^{-1}$, 
where $\Gamma$ is the bulk Lorentz factor of the jet plasma and $\beta=v/c$ its bulk velocity, depends on the viewing angle $\theta$, the flux at a given wavelength will be Doppler-enhanced according to the orientation of the corresponding emitting zone.
In the bottom panel of Fig.\ \ref{gamma}, we show the variation in the Doppler factors affecting the optical and mm fluxes under the hypothesis that their long-term trend is due to variations in the viewing angle alone and that the observed flux is proportional to $\delta^3$ \citep[see e.g.][]{urr95,vil99}, so that $\delta (t) =\delta_{\rm min}[F(t)/F_{\rm min}]^{1/3}$. 
We set $\delta_{\rm min}=11$, in the range of the values estimated by \citet{abd09_3c454} and \citet{ack10} to avoid pair production, hence self-absorption, of $\gamma$-ray radiation.
To reproduce the long-term trends we tentatively used cubic spline interpolations through the 7-d binned optical light curve and 30-d binned mm light curve\footnote{The mm emitting region is expected to be much wider than the optical one, so that its variability time scales would be longer.}.
In the bottom panel of Fig.\ \ref{gamma} we also show the evolution of the corresponding viewing angles, obtained as 
$\theta(t)=\rm \arccos \{ [ \Gamma \, \delta (t) -1]/[ \sqrt{\Gamma^2-1} \, \delta (t)] \}$,
where we put $\Gamma=15.6$, in agreement with the value derived by long-term Very Long Baseline Interferometry (VLBI) monitoring \citep{jor05}.
This produces maximum Doppler factors of $\sim 23.7$ for the optical and $\sim 22.5$ for the mm, and corresponding minimum viewing angles of about $2.1\degr$ and $2.3\degr$, respectively, at the time of flux maxima.
The minimum Doppler factor of 11, corresponding to a maximum angle of about $5.0 \degr$, was reached during the flux minima.
The above picture would explain the long/mid-term flux variations \citep[see also][]{vil02,vil04a}, while the short-term flares (see below) would be due either to perturbations moving downstream in the jet or to a finer geometrical structure, as proposed for BL Lacertae by \citet{lar10}.

As mentioned in the Introduction, the high-energy emission is commonly believed to come from inverse-Comptonisation of low-energy radiation according to either an SSC or an EC process, or a combination of both. In an SSC scenario, the dependence of the synchrotron emissivity on the density of relativistic particles is linear, while the dependence of the self-Compton emissivity is quadratic. 
However, that the increase of the flux levels of the $\gamma$ and X-ray light curves in the second observing season does not correspond to a rise of the optical and mm flux levels seems to rule out a variation in the electron density. The most straightward explanation is therefore that there has been a growth of the number of Comptonised seed photons, i.e.\ of the Comptonisation rate.

We investigate the matter in an SSC framework 
and we roughly divide the emitting jet in three zones,
where radiation at different frequencies is produced through the synchrotron process: the UV to near-IR region, which we will call the ``optical" region, the IR region, and the mm region\footnote{Indeed, in our inhomogeneous jet model the synchrotron emission at each frequency mainly comes from the jet region where that frequency is close to self absorption.}.
We assume that the variation in time of the number of both synchrotron (seed) photons and target electrons in the various zones is represented by the $R$-band light in the first region, by a linear combination of the $R$-band and mm light curves in the second region, and by the mm flux density variations in the third region.
We therefore try to fit the $\gamma$ and X-ray light curves in Fig.\ \ref{gamma} with the following general expression:
$$[s_R \, F_R(t)/\delta_R^3(t) + s_{\rm mm} \,F_{\rm mm}(t)/\delta_{\rm mm}^3(t)] \times [e_R \, F_R(t) + e_{\rm mm} \, F_{\rm mm}(t)]$$ 
where the first factor refers to the seed photons in the jet rest frame, the second factor to the Comptonising electrons, $F_R(t)$ and $F_{\rm mm}(t)$ represent cubic spline interpolations through the 1-d binned $R$-band and 7-d binned mm light curves, respectively, $\delta_R(t)$ and $\delta_{\rm mm}(t)$ are the Doppler factors affecting the optical and mm emissions, respectively, which are plotted in the bottom panel of Fig.\ \ref{gamma}, and $s_R, s_{\rm mm}, e_R, e_{\rm mm}$ are coefficients that regulate the relative contributions of the jet zones. 
The functions $F'_R(t)=F_R(t)/\delta_R^3(t)$ and $F'_{\rm mm}(t)=F_{\rm mm}(t)/\delta_{\rm mm}^3(t)$ represent the intrinsic optical and mm flux density changes, respectively, since $\delta^{-3}$ corrects for the variability due to the Doppler boosting. To better clarify this point, we superposed, on the optical and mm light curves in Fig.\ \ref{gamma}, those intrinsic changes that would be observed in case of a constant Doppler factor $\delta=18$, i.e.\ if there were no viewing angle variations. 
As one can see, ascribing the long-term trends to viewing angle variations on the chosen time scales (one week for the optical, one month for the mm) reveals intrinsic variations of similar amplitude, which in turn supports the initial assumption. 

The fit to the $\gamma$-ray light curve shown in the top panel of Fig.\ \ref{gamma} is obtained by setting $s_{\rm mm}=e_{\rm mm}=0$, suggesting that the $\gamma$-ray emission comes from the Comptonisation of synchrotron optical photons in the optical emitting region. However, due to the photon's mean free path before upscattering and to the fact that the synchrotron radiation is relativistically beamed forward, the bulk of the $\gamma$-ray radiation is expected to come from a region slightly downstream from the region responsible for the bulk of the optical emission.
To fit the X-ray light curve, we set $s_R=e_R$ and $s_{\rm mm}=e_{\rm mm}$, meaning that the X-ray emission is obtained by upscattering of IR photons on their parent relativistic electrons.
Also in this case we can imagine that the bulk of the X-ray radiation is produced a bit downstream from the bulk of the IR radiation.

The increase in the $\gamma$ and X-ray flux levels from the first to the second observing season can thus be explained as a growth in the number of seed photons that are Comptonised.
In particular, to explain the higher $\gamma$ flux in the second season, we need to increase the number of Comptonising interactions by about 60\%, while we need to increase it by a factor 1.6--1.8 to explain the higher X-ray flux.
This may be due to a variable misalignment between the zone responsible for the bulk of the synchrotron emission and the one responsible for the bulk of the respective inverse-Compton one.
In the second observing season, the regions would be better aligned, so that more seed photons can enter the corresponding Comptonisation zone.

In summary, we imagine a scenario where the variation in the angle between our line of sight and the optical emitting region in the 2008--2009 observing season is similar to the variation in the following 2009--2010 season, implying similar flux density levels. The same is true for the mm region. At the same time, the $\gamma$ and X-ray production zones become more aligned with the optical and IR regions in the second period, explaining the increased $\gamma$ and X-ray flux levels. 

\subsection{Short-term correlation between the optical and $\gamma$-ray fluxes}

The correlation between the $\gamma$-ray and optical fluxes of 3C 454.3 has been analysed in a number of papers including AGILE or Fermi data.
Using AGILE and WEBT data, \citet{ver09} find a mild correlation with no evident time delay in November 2007, while the increased optical activity in December allowed \citet{don09b} to constrain the $\gamma$-optical time lag to $-0.6^{+0.7}_{-0.5} \, \rm d$, indicating that the $\gamma$-ray flux variations follow the optical ones by $\sim 0$--1 d, even if a very fast $\gamma$ and optical flare on December 12 seemed to constrain the delay within 12 h.
When considering 18 months of AGILE and GASP-WEBT observations (2007 July -- 2009 January), \citet{ver10} found a time lag of $-0.4^{+0.6}_{-0.8}$ d of the $\gamma$-ray variations after the optical ones, consistent with the earlier results.
The cross-correlation study performed by \citet{bon09} on $\gamma$-ray data from Fermi and optical/IR data from SMARTS in 2008 found no detectable lag between IR/optical and $\gamma$-ray fluxes.

\begin{figure}
   \centering
   \resizebox{\hsize}{!}{\includegraphics{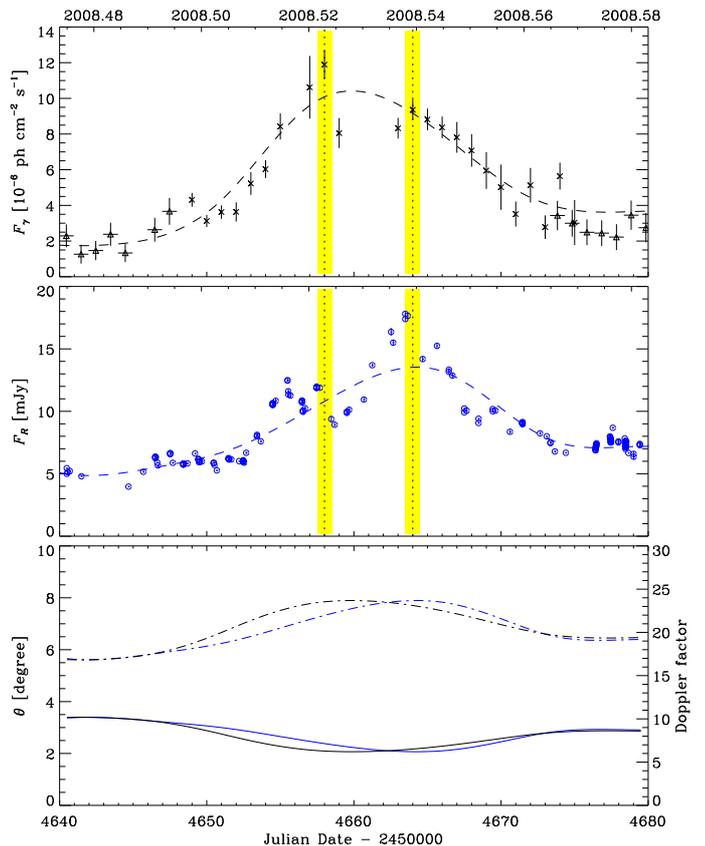}} 
   \caption{Flux variations of 3C 454.3 at $\gamma$-ray (top) and optical (middle) frequencies in mid 2008. The $\gamma$-ray light curve is built with Fermi data (crosses) from \citet{abd09_3c454} and AGILE data (triangles) from \citet{ver10}. Vertical dotted lines indicate the position of the two $\gamma$ peaks, the yellow stripes highlighting the data integration interval.
Cubic spline interpolations through the 7-day binned optical (blue dashed line) and $\gamma$ (black dashed line) light curves are also shown. They represent the variations due to geometrical changes.
The evolution of the Doppler factor $\delta$ (dot-dashed lines) and viewing angle $\theta$ (solid lines) is shown in the bottom panel for both the $\gamma$-ray (black) and optical (blue) bands.}
    \label{gamma_p1}
    \end{figure}

Figure \ref{gamma_p1} shows an enlargement of the $\gamma$ and optical light curves of Fig.\ \ref{gamma} around the brightest phase of the 2008 outburst.
The $\gamma$-ray flare appears double-peaked, the flux of the first maximum being 27\% higher than that of the second one.
The shape of the optical flare seems more complex, but we notice that there is an optical minor peak corresponding to the first $\gamma$ maximum, just before or simultaneous to it, and there is a major optical peak just before or simultaneous to the second $\gamma$ peak.
It is not clear whether there are really more optical events than $\gamma$ ones, or if other $\gamma$ peaks are hidden by insufficient sampling and precision. 
Moreover, it is not easy to understand why a minor (major) optical flare corresponds to a major (minor) event at $\gamma$ energies. 

\begin{figure}
   \centering
   \resizebox{\hsize}{!}{\includegraphics{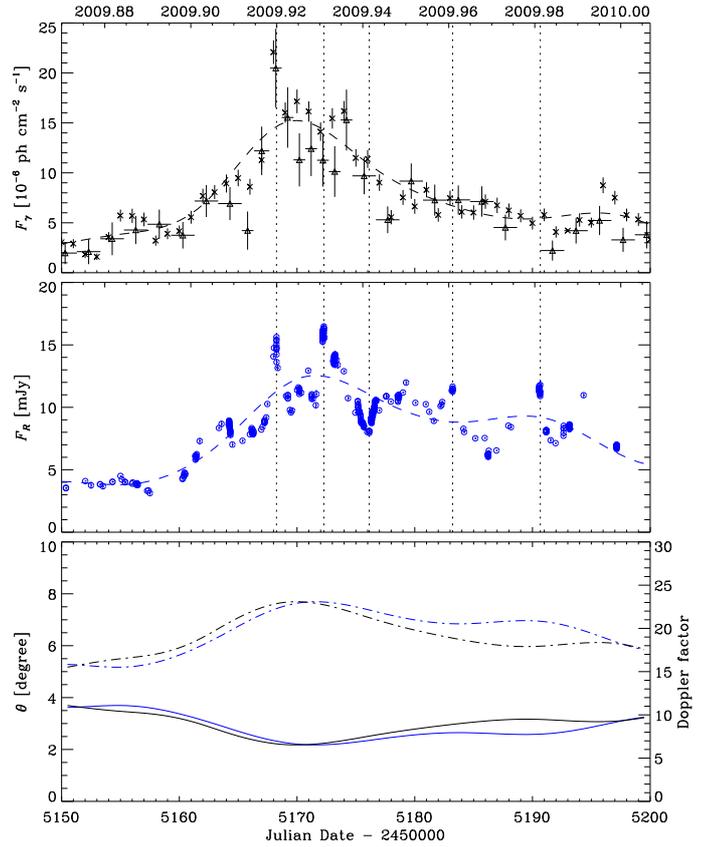}} 
   \caption{Flux variations of 3C 454.3 at $\gamma$-ray (top) and optical (middle) frequencies in late 2009. The $\gamma$-ray light curve is built with Fermi data (crosses) from \citet{ack10} and AGILE data (triangles) from \citet{pac10}. Vertical dotted lines indicate remarkable optical events. 
Cubic spline interpolations through the 7-day binned optical (blue dashed line) and $\gamma$ (black dashed line) light curves are also shown. They represent the variations due to geometrical changes.
The evolution of the Doppler factor $\delta$ (dot-dashed lines) and viewing angle $\theta$ (solid lines) is shown in the bottom panel for both the $\gamma$-ray (black) and optical (blue) bands.}
    \label{gamma_p2}
    \end{figure}

The period of most intense activity in late 2009 is shown in Fig.\ \ref{gamma_p2}.
We can recognise some features similar to those noticed in the light curves of mid 2008.
The dense optical monitoring performed by the GASP collaboration allowed us to detect a sharp peak on $\rm JD = 2455168$, simultaneous to the major $\gamma$-ray one. 
Then, a second major fast optical flare at $\rm JD = 2455172$ found the $\gamma$-ray flux at a much lower level, possibly in between two little $\gamma$ bumps. 
We again notice a difference in the flux ratios: the second optical peak is brighter than the first one, whereas the opposite is true in $\gamma$-rays.
Moreover, a well-sampled optical minimum is visible at $\rm JD = 2455176$, while the minimum is reached 1--2 days later at $\gamma$ frequencies. 
Finally, the optical flares peaking at $\rm JD = 2455183$ and $\rm JD = 2455190$ have no evident $\gamma$-ray counterparts.

In summary, when inspecting the $\gamma$-optical correlation on short time scales,
sometimes the optical and $\gamma$ flux variations are simultaneous, whereas in other cases the latter seem to be lacking or to possibly be delayed.

In the geometrical scenario proposed above to explain the long-term flux variations of the optical and mm light curves, as well as the increased $\gamma$ and X-ray fluxes in 2009--2010, a slight misalignment between the regions responsible for the production of the optical and $\gamma$-ray radiations can also account for the variation in the $\gamma$/optical flux ratio. In the bottom panels of Figs.\ \ref{gamma_p1} and \ref{gamma_p2} we show the evolution of the Doppler factors $\delta$ and viewing angles $\theta$ affecting the optical and $\gamma$ fluxes in the considered periods. 
The trends of the optical $\delta$ and $\theta$ have already been presented in the bottom panel of Fig.\ \ref{gamma}. To obtain the $\gamma$-ray ones we used cubic spline interpolations through the 7-d binned Fermi light curve and rescaled it to the optical one to take into account that we are comparing fluxes deriving from different processes. We therefore constructed a new $\gamma$-ray spline as $F_{\rm res}=a \, F^b$, which has the same flux maximum and minimum as the optical one. We got $a=3.53$ and $b=0.57$ in the period shown in Fig.\ \ref{gamma_p1}, while $a=1.70$ and $b=0.73$ in the period plotted in Fig.\ \ref{gamma_p2}\footnote{This means that in the considered periods the total long-term variation in the $\gamma$ flux is less than quadratic with respect to the optical one, as expected, because the quadratic dependence of the intrinsic fluxes is ``diluted" by the Doppler factor variation.}.
In both periods one can see that the $\gamma$-emitting region acquires the minimum $\theta$, corresponding to the maximum $\delta$, slightly before than the optical zone.
We suggest that in both periods, because of the different orientation of the emitting regions, the $\gamma$-ray flux was more relativistically boosted than the optical one at the time of the first $\gamma$ peak, while the opposite occurred at the time of the second peak. 
Moreover, we can see that, in the rising part of the 2009 outburst, the two regions are better aligned, yielding more comparable optical and $\gamma$ flares than in the rising part of the 2008 outburst. 
The maximum misalignment between the optical and $\gamma$ emitting regions is quite small: $\sim 0.4 \degr$ in mid 2008 and $\sim 0.6 \degr$ in late 2009.

\subsection{Short-term correlation between the X-ray and optical/$\gamma$ fluxes}

\begin{figure}
   \centering
   \resizebox{\hsize}{!}{\includegraphics{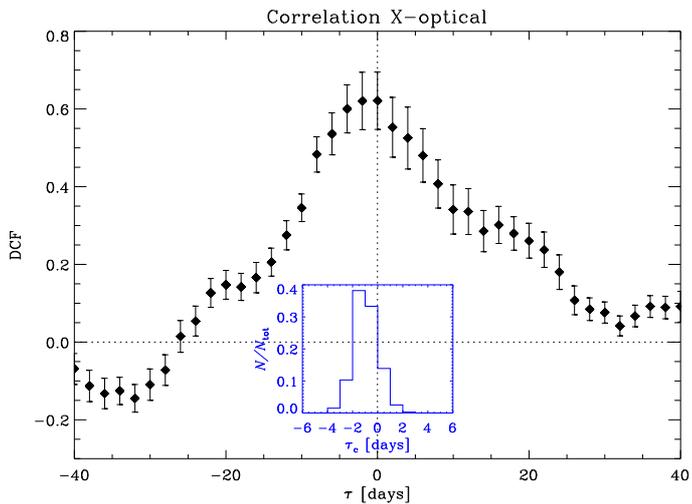}} 
   \caption{Discrete correlation function (DCF) between the 1 keV flux densities from XRT and the GASP $R$-band flux densities. The inset shows the results of 1000 Monte Carlo simulations.}
    \label{dcf_xo}
    \end{figure}

The X-ray light curve is less sampled than the $\gamma$-ray one, so that a detailed analysis of correlation like the one performed in the previous section is not possible.
We thus investigate the optical-X correlation by analysing the entire long-term light curves shown in Fig.\ \ref{gamma} by means of the discrete correlation function \citep[DCF;][]{ede88,huf92}, which was specifically designed to study unevenly sampled data sets.
In Fig.\ \ref{dcf_xo} we show the DCF obtained by cross-correlating the 1 keV with the $R$-band flux densities. 
Both light curves have previously been binned over 12 hours, while the DCF bin was set to two days.
The curve peaks at a time lag $\tau$ between $-2$ and 0 d, with a DCF value of 0.62, which indicates mild correlation, and flux variations in the optical band leading or being simultaneous to those in the X-ray one. 
In general, a more precise measure of the time lag can be obtained by calculating the centroid of the DCF,
$\tau_{\rm c}=(\sum_i \tau_i {\rm DCF}_i)/(\sum_i {\rm DCF}_i)$, where sums
run over the points which have a DCF value close to the peak one.
In our case the calculation of the centroid using all points with $\rm DCF > 0.8 \, DCF_{peak}$ (6 points) yields $\tau_{\rm c}=-1.0 \, \rm d$, i.e.\ an X-ray lag of 1 d. 
The same result is obtained by lowering the threshold for the centroid calculation to 0.75 and 0.70 of the $\rm DCF_{peak}$ (8 points).
To estimate the uncertainty on this time lag, we performed Monte Carlo simulations following the technique known as ``flux redistribution/random subset selection" \citep[FR/RSS;][see also \citealt{rai03}]{pet98}, which allows testing the influence of both sampling and
errors on the results. We randomly selected subsets of the
X-ray and optical light curves, discarding redundant points,
and adding to the fluxes random Gaussian deviates constrained by the flux errors.
The X-ray and optical subsets are then cross-correlated and the resulting DCF
centroids stored. A measure of the lag uncertainty can then be
derived from the centroid distribution, by considering the range of $\tau_{\rm c}$ values containing 68.27\%  of the realisations. 
The inset in Fig.\ \ref{dcf_xo} displays the $\tau_{\rm c}$ distribution obtained from 1000 Monte Carlo realisations: $\tau_{\rm c}$ is in the range $-2$--0 d in 71\% of cases ($\ga 1 \sigma$). This allows us to conclude in a more precise way that the time lag of the X-ray flux variations after the optical ones is $\tau=1.0 \pm 1.0 \, \rm d$. 

The finding of a time delay of the X-ray variations after the optical ones agrees with the picture proposed above, where the X-ray radiation is produced in the IR region, downstream of the optical one.
In the proposed scenario we would also expect the X-ray emission to be slightly delayed with respect to the $\gamma$-ray one, which is supposed to be very close to the optical one. 
Cross-correlation of the $\gamma$-ray light curve with the X-ray one, with the same choice of binning parameters as in the X-ray/optical case, yields a highly significant peak ($\rm DCF_{peak} \sim 1$) at zero time lag, while $\tau_{\rm c}=1.0 \, \rm d$.
The result of the Monte Carlo method described above is $\tau = 0.5 \pm 1.0 \, \rm d$, confirming a possible slight delay in the X-ray flux variations. 

\section{Conclusions}

In this paper we have presented multifrequency observations of 3C 454.3 from April 2008 to March 2010. Radio-to-optical data are mainly from the GASP-WEBT, and UV and X-ray data are from Swift and $\gamma$-ray data from AGILE and Fermi. In these two observing seasons, the source showed prominent outbursts at frequencies greater than 8 GHz, with rapid flares superposed in the optical-to-$\gamma$ light curves.

We have reanalysed the question of the nature of the optical--UV emission, deriving more accurate calibrations for both the UVOT instrument onboard Swift and the OM detector onboard XMM-Newton, and estimating the contribution of the emission lines (mainly Ly$\alpha$) in the UV. We confirmed that the observed decrease of the variability amplitude with increasing frequency from the near-IR to the UV band is due to the contribution of radiation from both the BLR and the accretion disc.
The long-term flux variations from the optical to the $\gamma$-ray frequencies appear roughly correlated. 
The mm flux remained high during the whole optical flaring period.

By comparing the multiwavelength behaviour in the 2008--2009 and 2009--2010 observing seasons, we noticed that the optical flux reached nearly the same levels (i.e.\ minimum and maximum brightness) in both seasons. The same thing occurred in the mm band, whereas the $\gamma$-ray and X-ray flux levels increased in the second period. 

We interpreted the observations in terms of an inhomogeneous jet model, where synchrotron radiation of increasing wavelength is produced in progressively larger regions farther away from the emitting jet apex. The jet is supposed to be a curved (possibly in a helical shape) and flexible (maybe rotating) structure, where different emitting zones have different alignments with the line of sight, and their viewing angle can change in time. A similar model has been adopted to explain the multiwavelength behaviour of Mkn 501 \citep{vil99}, S4 0954+65 \citep{rai99}, BL Lacertae \citep{vil02,vil04a,vil09a,rai09,rai10,lar10}, AO 0235+164 \citep{ost04}, along with 3C 454.3 \citep{vil06,vil07,vil09b}.

We assumed that the long-term flux density variations, with a tentative time scale of one week in the optical and of one month in the mm band, are due to changes of the viewing angles of the corresponding emitting regions, while variability on shorter time scales can be ascribed either to intrinsic flaring activity or to a finer geometrical structure (as suggested by \citealt{lar10} for BL Lacertae).
Then, under the further assumption that the $\gamma$ and X-ray emissions are produced by inverse-Compton scattering of synchrotron photons and that the behaviour of the IR light curve is intermediate between the optical and mm ones, we combined the optical and mm fluxes to fit the $\gamma$ and X-ray light curves.
In this scenario the multiwavelength observations presented in this paper suggest that the $\gamma$-ray radiation comes from inverse-Comptonisation of ``optical" (actually UV to-near-IR) synchrotron photons produced in the inner emitting jet region by their parent relativistic electrons (SSC process).
However, depending on the mean free path before upscattering, the bulk of the $\gamma$-ray radiation is produced in a region slightly downstream of the one responsible for the bulk of the optical emission. 
The jet curvature then implies that the $\gamma$ and optical regions are slightly misaligned.
A reduction of this misalignment, allowing more optical seed photons to enter the Comptonisation zone, could thus be at the origin of the $\gamma$-ray flux increase in 2009--2010.
Moreover, the variation in the $\gamma$/optical flux ratio during the outburst peaks of mid 2008 and late 2009 can be explained by the fact that the $\gamma$ zone acquires the minimum viewing angle, hence the maximum Doppler boosting, some days before the optical zone.

The X-ray light curve is not as well sampled as the optical and $\gamma$-ray ones, which makes a detailed analysis more difficult. 
However, the above-mentioned fit suggests that the X-ray radiation comes from a downstream region, where relativistic electrons produce and upscatter synchrotron IR photons (SSC process). 
The rise in the X-ray flux levels in the 2009--2010 season would then be explained similarly to the case of the $\gamma$ flux, with a better alignment between the zones responsible for the bulk of the synchrotron and inverse-Compton emissions.
The mean delay of about 1 d of the X-ray flux variations after the optical ones and of about 0.5 d after the $\gamma$-ray ones, found by cross-correlation analysis, are consistent with this picture.

\begin{acknowledgements}
C.\ M.\ R.\ is grateful to Stefania Rasetti from the Torino Observatory Computer Centre for her assistance.
We thank the Fermi-LAT team for providing Fermi data.
We acknowledge the use of public data from the Swift data archive.
We acknowledge financial contribution from the agreement ASI-INAF I/009/10/0.
St.-Petersburg University team acknowledges support from Russian RFBR foundation via grant 09-02-00092.
Abastumani Observatory team acknowledges financial support by the Georgian National Science Foundation through grant GNSF/ST08/4-404.
The Mets\"ahovi team acknowledges the support from the Academy of Finland
to our observing projects (numbers 212656, 210338, 121148, and others).
The University of Michigan team acknowledges financial support through the 
NASA Fermi grants NNX09AU16G, NNX10AP16G, and NSF grant AST-0607523.
The operation of UMRAO is funded by the University of Michigan.
This paper is partly based on observations carried out at the German-Spanish Calar Alto Observatory, which is jointly operated by the MPIA and the IAA-CSIC. Acquisition of the MAPCAT data is supported in part by MICIIN (Spain) grant and AYA2010-14844, and by CEIC (Andaluc\'{i}a) grant P09-FQM-4784.
Partly based on observations with the Medicina and Noto telescopes operated by INAF - Istituto di Radioastronomia.
This research has made use of NASA's Astrophysics Data System and of the XRT Data Analysis Software (XRTDAS) developed under the responsibility of the ASI Science Data Center (ASDC), Italy.
This research has made use of data from the MOJAVE database that is maintained by the MOJAVE team \citep{lis09}.
The Submillimeter Array is a joint project between the Smithsonian Astrophysical Observatory and the Academia Sinica Institute of Astronomy and Astrophysics and is funded by the Smithsonian Institution and the Academia Sinica.
The research at Boston U. was supported by NSF grant AST-0907893 and NASA Fermi GI grants NNX08AV65G, NNX08AV61G, and NNX09AT99G. 
The PRISM camera at Lowell Observatory was developed by K.\ Janes et al. at BU and Lowell Observatory, with funding from the NSF, BU, and Lowell Observatory.

\end{acknowledgements}

\longtab{3}{
\begin{longtable}{c c c c c c c c}
\caption{\label{xrt} Results of the analysis of the XRT observations}\\    
\hline\hline                 
   ObsID    &   START     & JD          & Exp  & Mode & $\Gamma$        & $F_{\rm 1 keV}$ & $\chi^2/\nu \, (\nu)$\\
            &             &$-2450000$   & [s]  &      &                 & [$\mu Jy$]      &                   \\
\hline 
\endfirsthead
\caption{continued.}\\
\hline\hline
   ObsID    &   START    & JD           & Exp  & Mode & $\Gamma$        & $F_{\rm 1 keV}$ & $\chi^2/\nu \, (\nu)$\\
            &            &$-2450000$    & [s]  &      &                 & [$\mu Jy$]      &                   \\
\hline
\endhead
\hline
\endfoot
00031216001 & 2008-05-27 16:45:01 & 4614.19793 & 3982 & PC*  & $1.54 \pm 0.07$ & $3.81 \pm 0.22$ & 0.95 (65) \\
00031216002 & 2008-05-28 04:08:01 & 4614.67223 & 2023 & PC*  & $1.57 \pm 0.08$ & $4.26 \pm 0.29$ & 1.16 (45) \\
00031216003 & 2008-05-30 12:18:00 & 4617.01250 & 2144 & PC*  & $1.56 \pm 0.07$ & $5.09 \pm 0.33$ & 0.99 (53) \\
00031216004 & 2008-05-31 12:22:01 & 4618.01529 & 2144 & PC*  & $1.47 \pm 0.06$ & $5.90 \pm 0.35$ & 1.12 (68) \\ 
00031216005 & 2008-06-03 18:45:10 & 4621.28137 & 1995 & PC*  & $1.57 \pm 0.10$ & $4.16 \pm 0.37$ & 1.19 (32) \\ 
00031216006 & 2008-06-04 18:50:57 & 4622.28538 & 1995 & PC*  & $1.56 \pm 0.09$ & $4.20 \pm 0.31$ & 0.68 (40) \\ 
00031216007 & 2008-06-05 01:35:00 & 4622.56597 & 2137 & PC*  & $1.59 \pm 0.08$ & $4.82 \pm 0.33$ & 0.66 (46) \\
00031216008 & 2008-06-06 00:19:01 & 4623.51321 & 2079 & PC*  & $1.59 \pm 0.09$ & $4.21 \pm 0.31$ & 0.71 (42) \\
00031216009 & 2008-06-07 02:03:01 & 4624.58543 & 1807 & PC*  & $1.59 \pm 0.09$ & $4.66 \pm 0.35$ & 1.29 (41) \\    
00031216010 & 2008-07-24 00:10:01 & 4671.50696 & 1952 & PC*  & $1.55 \pm 0.07$ & $5.68 \pm 0.35$ & 0.81 (57) \\   
00031216011 & 2008-07-26 14:43:01 & 4674.11321 & 1969 & PC*  & $1.47 \pm 0.06$ & $6.23 \pm 0.36$ & 1.31 (69) \\
00031216012 & 2008-07-29 00:32:00 & 4676.52222 & 1769 & WT   & $1.65 \pm 0.05$ & $7.07 \pm 0.30$ & 0.99 (117) \\         
00031216013 & 2008-07-30 11:52:00 & 4677.99444 & 1939 & PC*  & $1.49 \pm 0.07$ & $6.88 \pm 0.42$ & 0.81 (61) \\ 
00031216014 & 2008-08-01 00:25:01 & 4679.51737 & 1963 & PC*  & $1.54 \pm 0.07$ & $5.56 \pm 0.34$ & 0.98 (58) \\
00031216017 & 2008-08-01 11:37:30 & 4679.98438 & 1999 & WT   & $1.61 \pm 0.07$ & $7.77 \pm 0.44$ & 1.50 (84) \\  
00031216015 & 2008-08-03 10:34:01 & 4681.94029 & 2043 & PC*  & $1.54 \pm 0.08$ & $5.83 \pm 0.38$ & 0.82 (55) \\
00031216016 & 2008-08-04 07:19:01 & 4682.80487 & 1915 & PC*  & $1.53 \pm 0.09$ & $4.53 \pm 0.34$ & 1.23 (44) \\
00031216018 & 2008-08-05 21:41:01 & 4684.40348 &  380 & PC*  & $1.64 \pm 0.20$ & $5.59_{-0.82}^{+0.91}$ & Cash \\ 
00031216019 & 2008-08-06 07:19:01 & 4684.80487 & 2276 & PC*  & $1.60 \pm 0.08$ & $5.04 \pm 0.33$ & 1.02 (54) \\
00031216020 & 2008-08-07 20:44:01 & 4686.36390 & 1803 & PC*  & $1.61 \pm 0.09$ & $5.20 \pm 0.39$ & 1.02 (38) \\
00031216021 & 2008-08-08 12:47:01 & 4687.03265 & 4217 & PC*  & $1.64 \pm 0.06$ & $5.34 \pm 0.26$ & 1.17 (89) \\
00031216022 & 2008-08-10 01:47:01 & 4688.57432 & 2753 & PC*  & $1.62 \pm 0.09$ & $5.56 \pm 0.43$ & 1.08 (35) \\
00031216023 & 2008-08-12 06:32:01 & 4690.77223 & 2190 & PC*  & $1.55 \pm 0.07$ & $5.31 \pm 0.32$ & 1.18 (62) \\
00031216024 & 2008-08-14 04:52:00 & 4692.70278 & 1355 & PC*  & $1.52 \pm 0.13$ & $4.10 \pm 0.44$ & 1.38 (21) \\
00031216025 & 2008-08-15 01:47:01 & 4693.57432 & 1649 & PC*  & $1.63 \pm 0.10$ & $4.22 \pm 0.34$ & 0.98 (33) \\ 
00031216026 & 2008-08-16 21:07:01 & 4695.37987 & 1969 & PC*  & $1.59 \pm 0.08$ & $5.19 \pm 0.35$ & 1.04 (50) \\
00031216027 & 2008-08-17 16:43:00 & 4696.19653 & 2425 & PC*  & $1.61 \pm 0.07$ & $5.72 \pm 0.32$ & 0.80 (66) \\
00031216029 & 2008-08-20 08:59:01 & 4698.87432 & 1846 & PC*  & $1.40 \pm 0.07$ & $4.72 \pm 0.33$ & 0.90 (51) \\
00031216030 & 2008-08-21 02:20:01 & 4699.59723 & 2095 & PC*  & $1.61 \pm 0.08$ & $4.90 \pm 0.34$ & 1.08 (44) \\
00031216031 & 2008-08-22 10:28:01 & 4700.93612 & 1975 & PC*  & $1.54 \pm 0.08$ & $4.64 \pm 0.32$ & 1.13 (46) \\
00031216032 & 2008-08-23 08:58:01 & 4701.87362 & 2155 & PC*  & $1.57 \pm 0.07$ & $5.68 \pm 0.34$ & 1.41 (63) \\
00031216033 & 2008-08-24 14:20:01 & 4703.09723 &  538 & PC*  & $1.58_{-0.20}^{+0.21}$ & $4.97_{-0.76}^{+0.78}$ & 1.30 (11) \\
00031216034 & 2008-08-25 09:37:01 & 4703.90071 &  335 & PC*  & $1.52 \pm 0.18$ & $6.20_{-0.89}^{+0.97}$ & Cash \\   
00031216035 & 2008-08-26 20:45:00 & 4705.36458 & 1059 & PC*  & $1.55 \pm 0.12$ & $4.91 \pm 0.50$ & 0.81 (23) \\
00031216036 & 2008-08-27 17:39:00 & 4706.23542 & 1957 & PC*  & $1.53 \pm 0.08$ & $4.78 \pm 0.35$ & 0.86 (43) \\
00031216038 & 2008-08-29 19:22:01 & 4708.30696 & 1885 & PC*  & $1.46 \pm 0.09$ & $3.99 \pm 0.33$ & 1.12 (37) \\
00031216039 & 2008-08-31 14:36:01 & 4710.10834 & 1597 & PC*  & $1.51 \pm 0.09$ & $4.39 \pm 0.34$ & 1.19 (37) \\
00031216040 & 2008-09-01 19:41:01 & 4711.32015 & 1730 & PC*  & $1.54 \pm 0.09$ & $4.64 \pm 0.37$ & 0.78 (37) \\
00031216041 & 2008-09-02 18:05:01 & 4712.25348 & 2863 & PC*  & $1.52 \pm 0.07$ & $4.05 \pm 0.26$ & 1.05 (59) \\
00031216042 & 2008-09-03 05:15:01 & 4712.71876 & 2052 & PC*  & $1.65 \pm 0.08$ & $4.59 \pm 0.31$ & 0.79 (44) \\   
00031216043 & 2008-09-04 16:54:46 & 4714.20470 &  346 & PC*  & $1.59 \pm 0.22$ & $4.02_{-0.68}^{+0.76}$ & Cash \\
00031216044 & 2008-09-05 01:00:01 & 4714.54168 & 1711 & PC*  & $1.62 \pm 0.09$ & $4.83 \pm 0.36$ & 1.50 (42) \\
00031216046 & 2008-09-07 15:27:40 & 4717.14421 &  470 & PC*  & $1.54 \pm 0.16$ & $4.96_{-0.66}^{+0.72}$ & Cash \\
00031216047 & 2008-09-08 10:33:00 & 4717.93958 & 2501 & PC*  & $1.54 \pm 0.07$ & $5.03 \pm 0.32$ & 0.63 (52) \\
00090023001 & 2008-09-09 01:20:01 & 4718.55557 & 1815 & PC*  & $1.51 \pm 0.09$ & $4.73 \pm 0.37$ & 1.22 (38) \\
00031216048 & 2008-09-10 17:32:01 & 4720.23057 & 1046 & PC*  & $1.53 \pm 0.12$ & $6.44 \pm 0.66$ & 0.98 (23) \\
00031216049 & 2008-09-11 12:58:00 & 4721.04028 & 2001 & PC*  & $1.56 \pm 0.08$ & $5.32 \pm 0.35$ & 1.04 (53) \\
00031216050 & 2008-09-12 20:58:01 & 4722.37362 & 1973 & PC*  & $1.66 \pm 0.08$ & $5.80 \pm 0.39$ & 0.76 (47) \\
00031216051 & 2008-09-13 20:55:02 & 4723.37155 &  411 & PC*  & $1.38 \pm 0.16$ & $5.13_{-0.70}^{+0.77}$ & Cash \\
00031216052 & 2008-09-16 15:03:00 & 4726.12708 &  837 & PC*  & $1.51 \pm 0.13$ & $4.93 \pm 0.52$ & 0.98 (21) \\
00031216053 & 2008-09-17 16:29:01 & 4727.18682 &  894 & PC*  & $1.56 \pm 0.12$ & $4.98 \pm 0.52$ & 0.99 (22) \\
00031216054 & 2008-09-23 05:36:01 & 4732.73334 & 3103 & PC*  & $1.58 \pm 0.08$ & $3.93 \pm 0.24$ & 0.89 (61) \\
00031216055 & 2008-09-24 09:14:01 & 4733.88473 & 1387 & PC*  & $1.66 \pm 0.12$ & $4.74 \pm 0.44$ & 0.89 (26) \\
00031216056 & 2008-09-25 02:34:01 & 4734.60696 & 2786 & PC*  & $1.60 \pm 0.07$ & $4.43 \pm 0.27$ & 1.13 (58) \\
00031216057 & 2008-09-26 04:16:01 & 4735.67779 & 2725 & PC*  & $1.55 \pm 0.07$ & $4.11 \pm 0.25$ & 0.70 (57) \\
00031216058 & 2008-09-27 04:22:00 & 4736.68194 & 2725 & PC*  & $1.62 \pm 0.08$ & $3.75 \pm 0.25$ & 1.11 (48) \\
00031216059 & 2008-09-28 04:27:01 & 4737.68543 & 1797 & PC*  & $1.57 \pm 0.08$ & $5.23 \pm 0.35$ & 0.93 (46) \\
00031216060 & 2008-09-29 03:08:01 & 4738.63057 & 2759 & PC*  & $1.59 \pm 0.06$ & $5.81 \pm 0.32$ & 1.09 (73) \\
00031216061 & 2008-09-30 03:06:01 & 4739.62918 & 2594 & PC*  & $1.60 \pm 0.06$ & $5.92 \pm 0.32$ & 1.29 (71) \\
00031216062 & 2008-10-01 04:42:00 & 4740.69583 & 3108 & PC*  & $1.66 \pm 0.06$ & $6.00 \pm 0.29$ & 1.04 (86) \\
00031216063 & 2008-10-02 04:47:01 & 4741.69932 & 3107 & PC*  & $1.64 \pm 0.06$ & $5.12 \pm 0.27$ & 1.20 (75) \\  
00035030028 & 2008-10-05 13:06:01 & 4745.04584 & 1221 & PC*  & $1.62 \pm 0.13$ & $4.09 \pm 0.40$ & 0.61 (23) \\
00035030029 & 2008-10-10 20:26:01 & 4750.35140 &  807 & PC*  & $1.78 \pm 0.16$ & $3.32_{-0.38}^{+0.42}$ & Cash \\

00035030030 & 2008-10-26 20:25:01 & 4766.35071 &  428 & PC*  & $1.69_{-0.23}^{+0.24}$ & $2.72_{-0.47}^{+0.54}$ & Cash \\
00090023002 & 2008-12-25 04:53:00 & 4825.70347 & 1493 & PC   & $1.62 \pm 0.15$ & $1.39_{-0.17}^{+0.16}$ & 0.93 (16) \\
00090023003 & 2008-12-26 05:15:01 & 4826.71876 & 1522 & PC   & $1.54 \pm 0.13$ & $1.49_{-0.16}^{+0.17}$ & 0.66 (19) \\ 
00090023004 & 2008-12-27 16:31:01 & 4828.18821 & 1682 & PC   & $1.77 \pm 0.20$ & $0.99_{-0.14}^{+0.16}$ & Cash\\
00090023005 & 2008-12-29 00:27:01 & 4829.51876 & 1331 & PC   & $1.68 \pm 0.16$ & $1.85_{-0.23}^{+0.25}$ & Cash \\
00090023006 & 2008-12-30 05:24:00 & 4830.72500 & 1276 & PC   & $1.46 \pm 0.18$ & $1.10_{-0.17}^{+0.16}$ & 1.24 (12) \\ 
00090023007 & 2008-12-31 16:42:01 & 4832.19584 &  707 & PC   & $1.58_{-0.22}^{+0.21}$ & $1.37 \pm 0.23$ & Cash \\ 
00090023008 & 2009-01-01 12:01:00 & 4833.00069 & 1131 & PC   & $1.57 \pm 0.21$ & $1.28 \pm 0.21$ & 1.72 (10) \\ 
00035030031 & 2009-01-21 17:05:01 & 4853.21182 & 1436 & PC   & $1.59_{-0.18}^{+0.19}$  & $1.34_{-0.19}^{+0.22}$ & Cash \\
00035030032 & 2009-01-22 17:17:10 & 4854.22025 &  857 & PC   & $1.73_{-0.23}^{+0.25}$ & $1.24_{-0.21}^{+0.22}$ & Cash \\ 
00035030033 & 2009-01-23 06:10:01 & 4854.75696 &  634 & PC   & $1.70_{-0.25}^{+0.27}$ & $1.58_{-0.28}^{+0.31}$ & Cash \\  
00035030034 & 2009-01-24 15:46:01 & 4856.15696 &  485 & PC   & $1.80_{-0.24}^{+0.25}$ & $1.66_{-0.29}^{+0.31}$ & Cash \\ 
00035030035 & 2009-01-25 06:32:00 & 4856.77222 &  927 & PC   & $1.70 \pm 0.16$ & $1.44_{-0.18}^{+0.20}$ & Cash \\
00035030036 & 2009-01-26 08:14:01 & 4857.84307 &  924 & PC   & $1.78 \pm 0.19$ & $1.75_{-0.25}^{+0.24}$ & 1.26 (11) \\
00035030037 & 2009-01-27 11:32:01 & 4858.98057 &  991 & PC   & $1.67 \pm 0.18$ & $1.43_{-0.21}^{+0.22}$ & 0.96 (10) \\ 
00035030038 & 2009-05-23 03:25:01 & 4974.64237 & 4137 & PC*  & $1.56 \pm 0.07$ & $2.46 \pm 0.16$ & 1.05 (52) \\
00035030039 & 2009-05-24 22:38:01 & 4976.44307 & 3805 & PC*  & $1.62 \pm 0.07$ & $3.62 \pm 0.20$ & 0.92 (70) \\
00035030040 & 2009-05-26 16:30:00 & 4978.18750 & 4061 & PC*  & $1.58 \pm 0.06$ & $3.31 \pm 0.19$ & 1.17 (67) \\ 
00031018009 & 2009-08-12 07:42:01 & 5055.82084 &  837 & PC*  & $1.65 \pm 0.14$ & $4.45 \pm 0.50$ & 1.20 (16) \\ 
00090081001 & 2009-08-13 12:55:01 & 5057.03821 & 2101 & PC*  & $1.75 \pm 0.08$ & $5.09 \pm 0.33$ & 1.17 (46) \\
00035030041 & 2009-08-14 00:09:01 & 5057.50626 & 1033 & PC*  & $1.74 \pm 0.12$ & $4.58_{-0.46}^{+0.47}$  &  1.39 (20) \\
00035030042 & 2009-08-15 01:47:02 & 5058.57433 & 1961 & PC*  & $1.70 \pm 0.09$ & $4.88 \pm 0.34$ & 0.73 (41) \\
00035030043 & 2009-08-16 11:17:01 & 5059.97015 &  451 & PC*  & $1.85_{-0.18}^{+0.19}$  &  $4.73_{-0.63}^{+0.68}$  & Cash \\
00035030044 & 2009-08-17 09:55:00 & 5060.91319 & 1783 & PC*  & $1.70 \pm 0.10$ & $4.36 \pm 0.35$ & 0.71 (32) \\
00035030045 & 2009-08-18 00:19:01 & 5061.51321 & 2178 & PC*  & $1.54 \pm 0.08$ & $4.04 \pm 0.29$ & 1.06 (43) \\
00035030046 & 2009-08-19 08:27:01 & 5062.85209 & 1367 & PC*  & $1.76 \pm 0.12$ & $4.59 \pm 0.43$ & 0.68 (24) \\
00035030047 & 2009-08-19 13:30:00 & 5063.06250 &  812 & PC*  & $1.62 \pm 0.14$ & $3.96 \pm 0.49$ & 1.31 (15) \\
00035030048 & 2009-08-20 06:55:01 & 5063.78821 & 1923 & PC*  & $1.66 \pm 0.08$ & $4.70 \pm 0.34$ & 0.75 (40) \\
00035030049 & 2009-08-21 21:33:00 & 5065.39792 & 1096 & PC*  & $1.68 \pm 0.13$ & $4.33 \pm 0.45$ & 0.92 (21) \\
00035030050 & 2009-08-22 19:57:52 & 5066.33185 & 1826 & PC*  & $1.51 \pm 0.08$ & $5.71 \pm 0.38$ & 1.20 (49) \\
00035030051 & 2009-08-23 08:47:00 & 5066.86597 & 2066 & PC*  & $1.51 \pm 0.07$ & $6.70 \pm 0.40$ & 1.01 (63) \\
00035030052 & 2009-08-24 08:55:01 & 5067.87154 & 2021 & PC*  & $1.51 \pm 0.06$ & $6.11 \pm 0.36$ & 1.20 (63) \\
00035030053 & 2009-08-25 08:48:00 & 5068.86667 & 1517 & PC*  & $1.51 \pm 0.09$ & $7.02 \pm 0.51$ & 1.05 (47) \\
00035030054 & 2009-08-26 21:59:00 & 5070.41597 & 1559 & PC*  & $1.47 \pm 0.06$ & $9.22 \pm 0.53$ & 0.68 (71) \\
00035030055 & 2009-08-27 09:01:01 & 5070.87571 & 1573 & PC*  & $1.38 \pm 0.07$ & $8.25 \pm 0.53$ & 0.75 (64) \\
00035030056 & 2009-08-28 09:15:01 & 5071.88543 & 1637 & PC*  & $1.39 \pm 0.07$ & $8.85 \pm 0.54$ & 1.08 (67) \\
00035030057 & 2009-08-29 09:10:01 & 5072.88196 & 1692 & PC*  & $1.57 \pm 0.06$ & $9.06 \pm 0.48$ & 0.98 (75) \\
00035030058 & 2009-08-30 15:42:01 & 5074.15418 & 2601 & PC*  & $1.45 \pm 0.05$ & $7.80 \pm 0.36$ & 1.06 (102) \\
00035030059 & 2009-08-31 15:47:01 & 5075.15765 & 2074 & PC*  & $1.50 \pm 0.06$ & $8.56 \pm 0.43$ & 1.01 (91) \\
00035030060 & 2009-09-01 14:17:00 & 5076.09514 & 2003 & PC*  & $1.52 \pm 0.06$ & $7.82 \pm 0.40$ & 1.13 (82) \\
00035030061 & 2009-09-02 17:55:42 & 5077.24701 & 2124 & PC*  & $1.52 \pm 0.06$ & $8.66 \pm 0.43$ & 0.94 (91) \\
00035030062 & 2009-09-03 06:54:01 & 5077.78751 &  963 & PC*  & $1.55 \pm 0.09$ & $8.86 \pm 0.72$ & 0.91 (37) \\
00035030063 & 2009-09-06 19:51:01 & 5081.32709 & 1647 & PC*  & $1.63 \pm 0.07$ & $7.11 \pm 0.44$ & 0.86 (54) \\
00035030064 & 2009-09-07 18:23:01 & 5082.26598 & 1467 & PC*  & $1.44 \pm 0.09$ & $5.71 \pm 0.47$ & 1.14 (38) \\
00035030065 & 2009-09-08 23:15:01 & 5083.46876 & 1442 & PC*  & $1.45 \pm 0.08$ & $7.68 \pm 0.55$ & 0.93 (46) \\
00035030066 & 2009-09-09 00:56:01 & 5083.53890 & 2116 & PC*  & $1.49 \pm 0.08$ & $5.44 \pm 0.40$ & 0.84 (44) \\
00090081002 & 2009-09-13 20:33:47 & 5088.35679 & 1993 & PC*  & $1.55 \pm 0.06$ & $9.53 \pm 0.53$ & 1.22 (70) \\
00035030067 & 2009-09-16 22:15:01 & 5091.42709 & 1503 & PC*  & $1.61 \pm 0.06$ & $9.89 \pm 0.54$ & 1.04 (71) \\
00031493001 & 2009-09-17 04:29:43 & 5091.68730 & 1999 & WT   & $1.51 \pm 0.04$ &$11.38 \pm 0.36$ & 1.00 (211) \\    
00031493003 & 2009-09-18 03:15:00 & 5092.63542 & 2509 & PC*  & $1.58 \pm 0.05$ & $9.75 \pm 0.40$ & 1.16 (118) \\
00031493004 & 2009-09-19 13:06:00 & 5094.04583 & 2542 & PC*  & $1.71 \pm 0.05$ & $8.02 \pm 0.36$ & 1.32 (94) \\
00031493005 & 2009-09-20 17:59:00 & 5095.24931 & 2389 & WT2  & $1.65 \pm 0.04$ & $7.96 \pm 0.28$ & 1.08 (185) \\     
00031493007 & 2009-09-22 00:26:01 & 5096.51807 & 1819 & WT2  & $1.68 \pm 0.05$ & $9.11 \pm 0.35$ & 1.08 (157) \\    
00031493008 & 2009-09-23 19:51:01 & 5098.32709 & 1578 & WT   & $1.72 \pm 0.05$ & $8.36 \pm 0.34$ & 0.99 (116) \\  
00031493009 & 2009-09-24 13:12:01 & 5099.05001 & 2109 & WT3  & $1.66 \pm 0.05$ & $8.64 \pm 0.34$ & 1.19 (132) \\ 
00031493010 & 2009-09-25 06:52:00 & 5099.78611 & 2149 & WT3  & $1.62 \pm 0.05$ & $7.66 \pm 0.31$ & 1.22 (137) \\ 
00031493011 & 2009-09-26 06:58:00 & 5100.79028 & 2144 & WT3  & $1.64 \pm 0.05$ & $7.85 \pm 0.33$ & 1.00 (126) \\ 
00031493012 & 2009-10-06 03:08:00 & 5110.63056 & 3422 & WT6  & $1.58 \pm 0.03$ &$11.28 \pm 0.28$ & 1.15 (285) \\ 
00031493013 & 2009-10-08 14:55:01 & 5113.12154 & 3416 & WT5  & $1.63 \pm 0.03$ &$11.31 \pm 0.27$ & 1.12 (282) \\ 
00031493014 & 2009-10-09 21:28:00 & 5114.39444 & 2917 & WT5  & $1.58 \pm 0.03$ &$12.33 \pm 0.32$ & 1.06 (265) \\ 
00090077001 & 2009-10-12 20:05:02 & 5117.33683 & 1958 & WT4  & $1.64 \pm 0.04$ & $9.70 \pm 0.36$ & 0.95 (158) \\ 
00090077002 & 2009-10-13 00:32:01 & 5117.52223 & 1783 & WT4  & $1.63 \pm 0.04$ & $9.41 \pm 0.36$ & 1.17 (148) \\ 
00090077003 & 2009-10-14 00:41:00 & 5118.52847 & 1838 & WT5  & $1.66 \pm 0.04$ & $8.54 \pm 0.39$ & 1.01 (110) \\ 
00090077004 & 2009-10-15 00:43:01 & 5119.52987 & 1854 & WT3  & $1.65 \pm 0.04$ & $8.38 \pm 0.33$ & 0.87 (130) \\ 
00090077005 & 2009-10-16 04:02:01 & 5120.66807 & 2114 & WT3  & $1.65 \pm 0.04$ & $8.38 \pm 0.33$ & 0.87 (130) \\ 
00090077006 & 2009-10-17 00:55:01 & 5121.53821 & 1964 & WT3  & $1.67 \pm 0.04$ & $9.70 \pm 0.33$ & 1.16 (185) \\ 
00090077007 & 2009-10-18 11:03:01 & 5122.96043 & 2119 & WT2  & $1.64 \pm 0.05$ & $7.66 \pm 0.30$ & 1.22 (147) \\ 
00090077008 & 2009-10-19 02:44:00 & 5123.61389 &  539 & WT2  & $1.70 \pm 0.09$ & $7.60 \pm 0.58$ & 1.20 (41) \\     
00090077009 & 2009-10-20 06:21:01 & 5124.76459 & 1868 & WT4  & $1.62 \pm 0.05$ & $7.51 \pm 0.31$ & 0.96 (134) \\ 
00090077010 & 2009-10-21 15:55:01 & 5126.16321 &  495 & WT   & $1.68 \pm 0.09$ & $8.82 \pm 0.63$ & 1.16 (44) \\      
00090077011 & 2009-10-22 19:25:01 & 5127.30904 & 2054 & WT2  & $1.75 \pm 0.04$ &$11.27 \pm 0.35$ & 1.23 (191) \\ 
00090077012 & 2009-10-23 00:27:01 & 5127.51876 & 4922 & WT6  & $1.68 \pm 0.02$ &$11.27 \pm 0.23$ & 1.18 (318) \\ 
00090077013 & 2009-10-24 00:01:01 & 5128.50071 & 1638 & WT4  & $1.67 \pm 0.05$ & $8.41 \pm 0.34$ & 0.87 (125) \\ 
00090077014 & 2009-10-25 19:41:00 & 5130.32014 & 2124 & WT2  & $1.67 \pm 0.04$ & $8.15 \pm 0.30$ & 1.01 (155) \\ 
00035030068 & 2009-10-31 12:08:01 & 5136.00557 & 1429 & WT2  & $1.65 \pm 0.06$ &$ 7.25 \pm 0.34$ & 0.76 (99) \\    
00035030069 & 2009-11-04 17:38:02 & 5140.23475 &  801 & WT4  & $1.78 \pm 0.09$ &$ 7.48 \pm 0.48$ & 1.29 (57) \\ 
00035030070 & 2009-11-06 23:56:00 & 5142.49722 & 1015 & WT   & $1.64 \pm 0.08$ &$ 6.23 \pm 0.39$ & 0.95 (66) \\     
00035030071 & 2009-11-11 17:45:00 & 5147.23958 & 1080 & WT   & $1.78 \pm 0.08$ &$ 6.20 \pm 0.37$ & 1.00 (66) \\   
00035030072 & 2009-11-18 02:33:01 & 5153.60626 & 1144 & WT   & $1.62 \pm 0.07$ &$ 6.48 \pm 0.37$ & 0.79 (75) \\    
00035030073 & 2009-11-27 20:49:01 & 5163.36737 &  940 & WT   & $1.52 \pm 0.06$ &$ 8.87 \pm 0.48$ & 1.03 (88) \\      
00035030074 & 2009-12-01 16:35:01 & 5167.19098 & 1065 & WT   & $1.56 \pm 0.05$ &$13.71 \pm 0.54$ & 1.11 (138) \\     
00035030075 & 2009-12-02 00:28:01 & 5167.51946 & 1174 & WT   & $1.53 \pm 0.04$ &$15.30 \pm 0.53$ & 1.20 (166) \\      
00035030076 & 2009-12-03 15:22:01 & 5169.14029 &  985 & WT   & $1.56 \pm 0.05$ &$15.42 \pm 0.60$ & 1.10 (139) \\    
00035030077 & 2009-12-04 00:39:01 & 5169.52709 & 2759 & WT2  & $1.49 \pm 0.02$ &$16.88 \pm 0.37$ & 1.09 (328) \\ 
00035030078 & 2009-12-05 13:37:38 & 5171.06780 &  645 & WT   & $1.50 \pm 0.05$ &$13.83 \pm 0.67$ & 1.07 (92) \\     
00035030079 & 2009-12-06 00:54:01 & 5171.53751 & 1079 & WT2  & $1.62 \pm 0.05$ &$13.79 \pm 0.59$ & 1.00 (117) \\ 
00035030080 & 2009-12-07 23:23:01 & 5173.47432 & 1114 & WT   & $1.51 \pm 0.04$ &$13.73 \pm 0.52$ & 1.36 (145) \\      
00035030081 & 2009-12-08 21:53:01 & 5174.41182 & 1164 & WT2  & $1.59 \pm 0.04$ &$13.58 \pm 0.51$ & 1.16 (143) \\ 
00035030082 & 2009-12-09 23:35:01 & 5175.48265 & 1100 & WT   & $1.48 \pm 0.04$ &$12.43 \pm 0.50$ & 1.32 (130) \\     
00035030083 & 2009-12-10 14:17:01 & 5176.09515 &  990 & WT   & $1.60 \pm 0.06$ &$10.90 \pm 0.51$ & 0.92 (104) \\     
00035030084 & 2009-12-11 09:20:01 & 5176.88890 & 1369 & WT2  & $1.54 \pm 0.05$ & $9.68 \pm 0.40$ & 1.24 (126) \\      
00035030085 & 2009-12-12 03:05:01 & 5177.62848 &  879 & WT2  & $1.59 \pm 0.08$ & $8.07 \pm 0.57$ & 1.03 (57) \\           
00035030086 & 2009-12-12 23:59:01 & 5178.49932 & 1325 & WT2  & $1.57 \pm 0.06$ & $7.19 \pm 0.37$ & 0.87 (90) \\        
00035030087 & 2009-12-14 01:35:01 & 5179.56598 & 1344 & WT3  & $1.62 \pm 0.05$ & $8.00 \pm 0.37$ & 0.94 (106) \\     
00035030088 & 2009-12-15 03:18:01 & 5180.63751 &  924 & WT2  & $1.56 \pm 0.07$ & $7.81 \pm 0.50$ & 1.24 (68) \\ 
00035030089 & 2009-12-16 03:23:01 & 5181.64098 & 1194 & WT2  & $1.56 \pm 0.06$ & $7.37 \pm 0.40$ & 1.14 (93) \\ 
00035030090 & 2009-12-18 00:22:01 & 5183.51529 & 1374 & WT   & $1.59 \pm 0.06$ & $8.17 \pm 0.39$ & 0.86 (110) \\       
00035030091 & 2009-12-19 11:45:01 & 5184.98959 & 1255 & WT   & $1.73 \pm 0.08$ &$10.98 \pm 0.73$ & 0.86 (61) \\      
00035030093 & 2009-12-21 02:15:01 & 5186.59376 & 1104 & WT   & $1.52 \pm 0.06$ & $8.38 \pm 0.42$ & 1.49 (98) \\    
00035030094 & 2009-12-22 10:22:00 & 5187.93194 & 1110 & WT   & $1.61 \pm 0.06$ & $8.44 \pm 0.42$ & 1.09 (95) \\        
00035030095 & 2009-12-25 07:46:01 & 5190.82362 & 1220 & WT   & $1.62 \pm 0.09$ & $6.98 \pm 0.56$ & 1.08 (47) \\     
00035030096 & 2009-12-26 01:06:00 & 5191.54583 & 1154 & WT   & $1.48 \pm 0.08$ & $7.16 \pm 0.51$ & 1.08 (60) \\     
00035030097 & 2009-12-26 23:59:01 & 5192.49932 &  869 & WT   & $1.62 \pm 0.08$ & $8.12 \pm 0.53$ & 0.79 (64) \\    
00035030098 & 2009-12-28 01:37:01 & 5193.56737 & 1200 & WT   & $1.63 \pm 0.06$ & $7.73 \pm 0.41$ & 1.04 (88) \\      
00035030099 & 2010-01-01 20:56:01 & 5198.37223 & 1109 & WT2  & $1.67 \pm 0.06$ & $9.47 \pm 0.45$ & 1.02 (98) \\     
00035030100 & 2010-01-08 01:02:01 & 5204.54307 & 1065 & WT   & $1.60 \pm 0.08$ & $6.52 \pm 0.43$ & 0.94 (64) \\     
00035030102 & 2010-01-17 19:11:01 & 5214.29932 & 2394 & WT3  & $1.63 \pm 0.03$ &$10.68 \pm 0.31$ & 0.95 (221) \\    
\hline                                  
\end{longtable}
}


\begin{thebibliography}{67}
\expandafter\ifx\csname natexlab\endcsname\relax\def\natexlab#1{#1}\fi

\bibitem[{{Abdo} {et~al.}(2011){Abdo}, {Ackermann}, {Ajello}, {Allafort},
  {Baldini}, {Ballet}, {Barbiellini}, {Bastieri}, {Bellazzini}, {Berenji}, \&
  {Blandford}}]{abd11}
{Abdo}, A.~A., {Ackermann}, M., {Ajello}, M., {et~al.} 2011, \apjl, 733, L26

\bibitem[{{Abdo} {et~al.}(2010){Abdo}, {Ackermann}, {Ajello}, {Antolini},
  {Baldini}, {Ballet}, {Barbiellini}, {Baring}, {Bastieri}, \&
  {Bechtol}}]{abd10_suzaku}
{Abdo}, A.~A., {Ackermann}, M., {Ajello}, M., {et~al.} 2010, \apj, 716, 835

\bibitem[{{Abdo} {et~al.}(2009){Abdo}, {Ackermann}, {Ajello}, {Atwood},
  {Axelsson}, {Baldini}, {Ballet}, {Barbiellini}, {Bastieri}, \&
  {Battelino}}]{abd09_3c454}
{Abdo}, A.~A., {Ackermann}, M., {Ajello}, M., {et~al.} 2009, \apj, 699, 817

\bibitem[{{Ackermann} {et~al.}(2010){Ackermann}, {Ajello}, {Baldini}, {Ballet},
  {Barbiellini}, {Bastieri}, {Bechtol}, {Bellazzini}, {Berenji}, {Blandford},
  {Bonamente}, {Borgland}, {Bregeon}, {Brigida}, {Bruel}, {Buehler}, {Burnett},
  \& {Buson}}]{ack10}
{Ackermann}, M., {Ajello}, M., {Baldini}, L., {et~al.} 2010, \apj, 721, 1383

\bibitem[{{Anderhub} {et~al.}(2009){Anderhub}, {Antonelli}, {Antoranz},
  {Backes}, {Baixeras}, {Balestra}, {Barrio}, {Bartko}, {Bastieri}, \& {Becerra
  Gonz{\'a}lez}}]{and09}
{Anderhub}, H., {Antonelli}, L.~A., {Antoranz}, P., {et~al.} 2009, \aap, 498,
  83

\bibitem[{{Bahcall} {et~al.}(1993){Bahcall}, {Bergeron}, {Boksenberg},
  {Hartig}, {Jannuzi}, {Kirhakos}, {Sargent}, {Savage}, {Schneider},
  {Turnshek}, {Weymann}, \& {Wolfe}}]{bah93}
{Bahcall}, J.~N., {Bergeron}, J., {Boksenberg}, A., {et~al.} 1993, \apjs, 87, 1

\bibitem[{{Ben{\'{\i}}tez} {et~al.}(2010){Ben{\'{\i}}tez}, {Chavushyan},
  {Raiteri}, {Villata}, {Dultzin}, {Mart{\'{\i}}nez}, {P{\'e}rez-Camargo}, \&
  {Torrealba}}]{benitez09}
{Ben{\'{\i}}tez}, E., {Chavushyan}, V.~H., {Raiteri}, C.~M., {et~al.} 2010, in
  Astronomical Society of the Pacific Conference Series, Vol. 427, Accretion
  and Ejection in AGN: a Global View, ed. {L.~Maraschi, G.~Ghisellini, R.~Della
  Ceca, \& F.~Tavecchio}, 291

\bibitem[{{Bessell} {et~al.}(1998){Bessell}, {Castelli}, \& {Plez}}]{bes98}
{Bessell}, M.~S., {Castelli}, F., \& {Plez}, B. 1998, \aap, 333, 231

\bibitem[{{Bonning} {et~al.}(2009){Bonning}, {Bailyn}, {Urry}, {Buxton},
  {Fossati}, {Maraschi}, {Coppi}, {Scalzo}, {Isler}, \& {Kaptur}}]{bon09}
{Bonning}, E.~W., {Bailyn}, C., {Urry}, C.~M., {et~al.} 2009, \apjl, 697, L81

\bibitem[{{Bonnoli} {et~al.}(2011){Bonnoli}, {Ghisellini}, {Foschini},
  {Tavecchio}, \& {Ghirlanda}}]{bon11}
{Bonnoli}, G., {Ghisellini}, G., {Foschini}, L., {Tavecchio}, F., \&
  {Ghirlanda}, G. 2011, \mnras, 410, 368

\bibitem[{{Burrows} {et~al.}(2005){Burrows}, {Hill}, {Nousek}, {Kennea},
  {Wells}, {Osborne}, {Abbey}, \& {Beardmore et al.}}]{bur05}
{Burrows}, D.~N., {Hill}, J.~E., {Nousek}, J.~A., {et~al.} 2005, Space Science
  Reviews, 120, 165

\bibitem[{{Cardelli} {et~al.}(1989){Cardelli}, {Clayton}, \& {Mathis}}]{car89}
{Cardelli}, J.~A., {Clayton}, G.~C., \& {Mathis}, J.~S. 1989, \apj, 345, 245

\bibitem[{{Cash}(1979)}]{cas79}
{Cash}, W. 1979, \apj, 228, 939

\bibitem[{{Dermer} {et~al.}(2009){Dermer}, {Finke}, {Krug}, \&
  {B{\"o}ttcher}}]{der09}
{Dermer}, C.~D., {Finke}, J.~D., {Krug}, H., \& {B{\"o}ttcher}, M. 2009, \apj,
  692, 32

\bibitem[{{Donnarumma} {et~al.}(2009){Donnarumma}, {Pucella}, {Vittorini},
  {D'Ammando}, {Vercellone}, {Raiteri}, {Villata}, {Perri}, {Chen}, {Smart},
  {Kataoka}, {Kawai}, {Mori}, {Tosti}, {Impiombato}, {Takahashi}, {Sato},
  {Tavani}, {Bulgarelli}, {Chen}, {Giuliani}, {Longo}, {Pacciani}, {Argan},
  {Barbiellini}, {Boffelli}, {Caraveo}, {Cattaneo}, {Cocco}, {Contessi},
  {Costa}, {Del Monte}, {De Paris}, {Di Cocco}, {Evangelista}, {Feroci},
  {Ferrari}, {Fiorini}, {Froysland}, {Frutti}, {Fuschino}, {Galli}, {Gianotti},
  {Labanti}, {Lapshov}, {Lazzarotto}, {Lipari}, {Marisaldi}, {Mastropietro},
  {Mereghetti}, {Morelli}, {Moretti}, {Morselli}, {Pellizzoni}, {Perotti},
  {Piano}, {Picozza}, {Pilia}, {Porrovecchio}, {Prest}, {Rapisarda},
  {Rappoldi}, {Rubini}, {Sabatini}, {Scalise}, {Soffitta}, {Striani},
  {Trifoglio}, {Trois}, {Vallazza}, {Zambra}, {Zanello}, {Pittori},
  {Santolamazza}, {Verrecchia}, {Giommi}, {Antonelli}, {Colafrancesco}, \&
  {Salotti}}]{don09b}
{Donnarumma}, I., {Pucella}, G., {Vittorini}, V., {et~al.} 2009, \apj, 707,
  1115

\bibitem[{{Edelson} \& {Krolik}(1988)}]{ede88}
{Edelson}, R.~A. \& {Krolik}, J.~H. 1988, \apj, 333, 646

\bibitem[{{Evans} \& {Koratkar}(2004)}]{eva04}
{Evans}, I.~N. \& {Koratkar}, A.~P. 2004, \apjs, 150, 73

\bibitem[{{Finke} \& {Dermer}(2010)}]{fin10}
{Finke}, J.~D. \& {Dermer}, C.~D. 2010, \apjl, 714, L303

\bibitem[{{Foschini} {et~al.}(2010){Foschini}, {Tagliaferri}, {Ghisellini},
  {Ghirlanda}, {Tavecchio}, \& {Bonnoli}}]{fos10}
{Foschini}, L., {Tagliaferri}, G., {Ghisellini}, G., {et~al.} 2010, \mnras,
  408, 448

\bibitem[{{Fuhrmann} {et~al.}(2006){Fuhrmann}, {Cucchiara}, {Marchili},
  {Tosti}, {Nucciarelli}, {Ciprini}, {Molinari}, {Chincarini}, {Zerbi},
  {Covino}, {Pian}, {Meurs}, {Testa}, {Vitali}, {Antonelli}, {Conconi},
  {Cutispoto}, {Malaspina}, {Nicastro}, {Palazzi}, \& {Ward}}]{fuh06}
{Fuhrmann}, L., {Cucchiara}, A., {Marchili}, N., {et~al.} 2006, \aap, 445, L1

\bibitem[{{Gehrels} {et~al.}(2004){Gehrels}, {Chincarini}, {Giommi}, {Mason},
  {Nousek}, {Wells}, {White}, {Barthelmy}, {Burrows}, {Cominsky}, {Hurley},
  {Marshall}, {M{\'e}sz{\'a}ros}, {Roming}, {Angelini}, {Barbier}, {Belloni},
  {Campana}, {Caraveo}, {Chester}, {Citterio}, {Cline}, {Cropper}, {Cummings},
  {Dean}, {Feigelson}, {Fenimore}, {Frail}, {Fruchter}, {Garmire}, {Gendreau},
  {Ghisellini}, {Greiner}, {Hill}, {Hunsberger}, {Krimm}, {Kulkarni}, {Kumar},
  {Lebrun}, {Lloyd-Ronning}, {Markwardt}, {Mattson}, {Mushotzky}, {Norris},
  {Osborne}, {Paczynski}, {Palmer}, {Park}, {Parsons}, {Paul}, {Rees},
  {Reynolds}, {Rhoads}, {Sasseen}, {Schaefer}, {Short}, {Smale}, {Smith},
  {Stella}, {Tagliaferri}, {Takahashi}, {Tashiro}, {Townsley}, {Tueller},
  {Turner}, {Vietri}, {Voges}, {Ward}, {Willingale}, {Zerbi}, \&
  {Zhang}}]{geh04}
{Gehrels}, N., {Chincarini}, G., {Giommi}, P., {et~al.} 2004, \apj, 611, 1005

\bibitem[{{Ghisellini} {et~al.}(2007){Ghisellini}, {Foschini}, {Tavecchio}, \&
  {Pian}}]{ghi07}
{Ghisellini}, G., {Foschini}, L., {Tavecchio}, F., \& {Pian}, E. 2007, \mnras,
  382, L82

\bibitem[{{Ghisellini} \& {Maraschi}(1989)}]{ghi89}
{Ghisellini}, G. \& {Maraschi}, L. 1989, \apj, 340, 181

\bibitem[{{Giommi} {et~al.}(2006){Giommi}, {Blustin}, {Capalbi},
  {Colafrancesco}, {Cucchiara}, {Fuhrmann}, {Krimm}, {Marchili}, {Massaro},
  {Perri}, {Tagliaferri}, {Tosti}, {Tramacere}, {Burrows}, {Chincarini},
  {Falcone}, {Gehrels}, {Kennea}, \& {Sambruna}}]{gio06}
{Giommi}, P., {Blustin}, A.~J., {Capalbi}, M., {et~al.} 2006, \aap, 456, 911

\bibitem[{{Gurwell} {et~al.}(2007){Gurwell}, {Peck}, {Hostler}, {Darrah}, \&
  {Katz}}]{gur07}
{Gurwell}, M.~A., {Peck}, A.~B., {Hostler}, S.~R., {Darrah}, M.~R., \& {Katz},
  C.~A. 2007, in Astronomical Society of the Pacific Conference Series, Vol.
  375, From Z-Machines to ALMA: (Sub)Millimeter Spectroscopy of Galaxies, ed.
  A.~J. {Baker}, J.~{Glenn}, A.~I. {Harris}, J.~G. {Mangum}, \& M.~S. {Yun},
  234

\bibitem[{{Hufnagel} \& {Bregman}(1992)}]{huf92}
{Hufnagel}, B.~R. \& {Bregman}, J.~N. 1992, \apj, 386, 473

\bibitem[{{Jorstad} {et~al.}(2010){Jorstad}, {Marscher}, {Larionov}, {Agudo},
  {Smith}, {Gurwell}, {L{\"a}hteenm{\"a}ki}, {Tornikoski}, {Markowitz},
  {Arkharov}, {Blinov}, {Chatterjee}, {D'Arcangelo}, {Falcone}, {G{\'o}mez},
  {Hagen-Thorn}, {Jordan}, {Kimeridze}, {Konstantinova}, {Kopatskaya},
  {Kurtanidze}, {Larionova}, {Larionova}, {McHardy}, {Melnichuk},
  {Roca-Sogorb}, {Schmidt}, {Skiff}, {Taylor}, {Thum}, {Troitsky}, \&
  {Wiesemeyer}}]{jor10}
{Jorstad}, S.~G., {Marscher}, A.~P., {Larionov}, V.~M., {et~al.} 2010, \apj,
  715, 362

\bibitem[{{Jorstad} {et~al.}(2005){Jorstad}, {Marscher}, {Lister}, {Stirling},
  {Cawthorne}, {Gear}, {G{\'o}mez}, {Stevens}, {Smith}, {Forster}, \&
  {Robson}}]{jor05}
{Jorstad}, S.~G., {Marscher}, A.~P., {Lister}, M.~L., {et~al.} 2005, \aj, 130,
  1418

\bibitem[{{Larionov} {et~al.}(2010){Larionov}, {Villata}, \& {Raiteri}}]{lar10}
{Larionov}, V.~M., {Villata}, M., \& {Raiteri}, C.~M. 2010, \aap, 510, A93

\bibitem[{{Lister} {et~al.}(2009){Lister}, {Aller}, {Aller}, {Cohen}, {Homan},
  {Kadler}, {Kellermann}, {Kovalev}, {Ros}, {Savolainen}, {Zensus}, \&
  {Vermeulen}}]{lis09}
{Lister}, M.~L., {Aller}, H.~D., {Aller}, M.~F., {et~al.} 2009, \aj, 137, 3718

\bibitem[{{Maraschi} {et~al.}(1992){Maraschi}, {Ghisellini}, \&
  {Celotti}}]{mar92}
{Maraschi}, L., {Ghisellini}, G., \& {Celotti}, A. 1992, \apjl, 397, L5

\bibitem[{{Neugebauer} {et~al.}(1979){Neugebauer}, {Oke}, {Becklin}, \&
  {Matthews}}]{neu79}
{Neugebauer}, G., {Oke}, J.~B., {Becklin}, E.~E., \& {Matthews}, K. 1979, \apj,
  230, 79

\bibitem[{{Nousek} \& {Shue}(1989)}]{nou89}
{Nousek}, J.~A. \& {Shue}, D.~R. 1989, \apj, 342, 1207

\bibitem[{{Ogle} {et~al.}(2010){Ogle}, {Wehrle}, {Balonek}, \&
  {Gurwell}}]{ogl10}
{Ogle}, P.~M., {Wehrle}, A.~E., {Balonek}, T., \& {Gurwell}, M.~A. 2010,
  arXiv:1003.3642

\bibitem[{{Ostorero} {et~al.}(2004){Ostorero}, {Villata}, \& {Raiteri}}]{ost04}
{Ostorero}, L., {Villata}, M., \& {Raiteri}, C.~M. 2004, \aap, 419, 913

\bibitem[{{Pacciani} {et~al.}(2010){Pacciani}, {Vittorini}, {Tavani},
  {Fiocchi}, {Vercellone}, {D'Ammando}, {Sakamoto}, {Pian}, {Raiteri},
  {Villata}, {Sasada}, {Itoh}, {Yamanaka}, {Uemura}, {Striani}, {Fugazza},
  {Tiengo}, {Krimm}, {Stroh}, {Falcone}, {Curran}, {Sadun}, {Lahteenmaki},
  {Tornikoski}, {Aller}, {Aller}, {Lin}, {Larionov}, {Leto}, {Takalo},
  {Berdyugin}, {Gurwell}, {Bulgarelli}, {Chen}, {Donnarumma}, {Giuliani},
  {Longo}, {Pucella}, {Argan}, {Barbiellini}, {Caraveo}, {Cattaneo}, {Costa},
  {De Paris}, {Del Monte}, {Di Cocco}, {Evangelista}, {Ferrari}, {Feroci},
  {Fiorini}, {Fuschino}, {Galli}, {Gianotti}, {Labanti}, {Lapshov},
  {Lazzarotto}, {Lipari}, {Marisaldi}, {Mereghetti}, {Morelli}, {Moretti},
  {Morselli}, {Pellizzoni}, {Perotti}, {Piano}, {Picozza}, {Pilia}, {Prest},
  {Rapisarda}, {Rappoldi}, {Rubini}, {Sabatini}, {Soffitta}, {Trifoglio},
  {Trois}, {Vallazza}, {Zanello}, {Colafrancesco}, {Pittori}, {Verrecchia},
  {Santolamazza}, {Lucarelli}, {Giommi}, \& {Salotti}}]{pac10}
{Pacciani}, L., {Vittorini}, V., {Tavani}, M., {et~al.} 2010, \apjl, 716, L170

\bibitem[{{Peterson}(2001)}]{pet01}
{Peterson}, B.~M. 2001, in Advanced Lectures on the Starburst-AGN Connection,
  ed. I.~{Aretxaga}, D.~{Kunth}, \& R.~{M{\'u}jica} (Singapore: World
  Scientific), 3

\bibitem[{{Peterson} {et~al.}(1998){Peterson}, {Wanders}, {Horne}, {Collier},
  {Alexander}, {Kaspi}, \& {Maoz}}]{pet98}
{Peterson}, B.~M., {Wanders}, I., {Horne}, K., {et~al.} 1998, \pasp, 110, 660

\bibitem[{{Pian} {et~al.}(2006){Pian}, {Foschini}, {Beckmann}, {Soldi},
  {T{\"u}rler}, {Gehrels}, {Ghisellini}, {Giommi}, {Maraschi}, {Pursimo},
  {Raiteri}, {Tagliaferri}, {Tornikoski}, {Tosti}, {Treves}, {Villata}, {Barr},
  {Courvoisier}, {di Cocco}, {Hudec}, {Fuhrmann}, {Malaguti}, {Persic},
  {Tavecchio}, \& {Walter}}]{pia06}
{Pian}, E., {Foschini}, L., {Beckmann}, V., {et~al.} 2006, \aap, 449, L21

\bibitem[{{Poole} {et~al.}(2008){Poole}, {Breeveld}, {Page}, {Landsman},
  {Holland}, {Roming}, {Kuin}, {Brown}, {Gronwall}, {Hunsberger}, {Koch},
  {Mason}, {Schady}, {vanden Berk}, {Blustin}, {Boyd}, {Broos}, {Carter},
  {Chester}, {Cucchiara}, {Hancock}, {Huckle}, {Immler}, {Ivanushkina},
  {Kennedy}, {Marshall}, {Morgan}, {Pandey}, {de Pasquale}, {Smith}, \&
  {Still}}]{poo08}
{Poole}, T.~S., {Breeveld}, A.~A., {Page}, M.~J., {et~al.} 2008, \mnras, 383,
  627

\bibitem[{{Raiteri} {et~al.}(2010){Raiteri}, {Villata}, {Bruschini}, \&
  {Capetti}}]{rai10}
{Raiteri}, C.~M., {Villata}, M., {Bruschini}, L., \& {Capetti}, A. 2010, \aap

\bibitem[{{Raiteri} {et~al.}(2009){Raiteri}, {Villata}, {Capetti}, {Aller},
  {Bach}, {Calcidese}, {Gurwell}, {Larionov}, {Ohlert}, {Nilsson},
  {Strigachev}, {Agudo}, {Aller}, {Bachev}, {Ben{\'{\i}}tez}, {Berdyugin},
  {B{\"o}ttcher}, {Buemi}, {Buttiglione}, {Carosati}, {Charlot}, {Chen},
  {Dultzin}, {Forn{\'e}}, {Fuhrmann}, {G{\'o}mez}, {Gupta}, {Heidt}, {Hiriart},
  {Hsiao}, {Jel{\'{\i}}nek}, {Jorstad}, {Kimeridze}, {Konstantinova},
  {Kopatskaya}, {Kostov}, {Kurtanidze}, {L{\"a}hteenm{\"a}ki}, {Lanteri},
  {Larionova}, {Leto}, {Latev}, {Le Campion}, {Lee}, {Ligustri}, {Lindfors},
  {Marscher}, {Mihov}, {Nikolashvili}, {Nikolov}, {Ovcharov}, {Principe},
  {Pursimo}, {Ragozzine}, {Robb}, {Ros}, {Sadun}, {Sagar}, {Semkov}, {Sigua},
  {Smart}, {Sorcia}, {Takalo}, {Tornikoski}, {Trigilio}, {Uckert}, {Umana},
  {Valcheva}, \& {Volvach}}]{rai09}
{Raiteri}, C.~M., {Villata}, M., {Capetti}, A., {et~al.} 2009, \aap, 507, 769

\bibitem[{{Raiteri} {et~al.}(2008{\natexlab{a}}){Raiteri}, {Villata}, {Chen},
  {Hsiao}, {Kurtanidze}, {Nilsson}, {Larionov}, {Gurwell}, {Agudo}, {Aller},
  {Aller}, {Angelakis}, {Arkharov}, {Bach}, {B{\"o}ttcher}, {Buemi},
  {Calcidese}, {Charlot}, {D'Ammando}, {Donnarumma}, {Forn{\'e}}, {Frasca},
  {Fuhrmann}, {G{\'o}mez}, {Hagen-Thorn}, {Jorstad}, {Kimeridze}, {Krichbaum},
  {L{\"a}hteenm{\"a}ki}, {Lanteri}, {Latev}, {Le Campion}, {Lee}, {Leto},
  {Lin}, {Marchili}, {Marilli}, {Marscher}, {Nesci}, {Nieppola},
  {Nikolashvili}, {Ohlert}, {Ovcharov}, {Principe}, {Pursimo}, {Ragozzine},
  {Sadun}, {Sigua}, {Smart}, {Strigachev}, {Takalo}, {Tavani}, {Thum},
  {Tornikoski}, {Trigilio}, {Uckert}, {Umana}, {Valcheva}, {Vercellone},
  {Volvach}, \& {Wiesemeyer}}]{rai08b}
{Raiteri}, C.~M., {Villata}, M., {Chen}, W.~P., {et~al.} 2008{\natexlab{a}},
  \aap, 485, L17

\bibitem[{{Raiteri} {et~al.}(1998){Raiteri}, {Villata}, {Lanteri}, {Cavallone},
  \& {Sobrito}}]{rai98}
{Raiteri}, C.~M., {Villata}, M., {Lanteri}, L., {Cavallone}, M., \& {Sobrito},
  G. 1998, \aaps, 130, 495

\bibitem[{{Raiteri} {et~al.}(2008{\natexlab{b}}){Raiteri}, {Villata},
  {Larionov}, {Gurwell}, {Chen}, {Kurtanidze}, {Aller}, {B{\"o}ttcher},
  {Calcidese}, {Hroch}, {L{\"a}hteenm{\"a}ki}, {Lee}, {Nilsson}, {Ohlert},
  {Papadakis}, {Agudo}, {Aller}, {Angelakis}, {Arkharov}, {Bach}, {Bachev},
  {Berdyugin}, {Buemi}, {Carosati}, {Charlot}, {Chatzopoulos}, {Forn{\'e}},
  {Frasca}, {Fuhrmann}, {G{\'o}mez}, {Gupta}, {Hagen-Thorn}, {Hsiao}, {Jordan},
  {Jorstad}, {Konstantinova}, {Kopatskaya}, {Krichbaum}, {Lanteri},
  {Larionova}, {Latev}, {Le Campion}, {Leto}, {Lin}, {Marchili}, {Marilli},
  {Marscher}, {McBreen}, {Mihov}, {Nesci}, {Nicastro}, {Nikolashvili}, {Novak},
  {Ovcharov}, {Pian}, {Principe}, {Pursimo}, {Ragozzine}, {Ros}, {Sadun},
  {Sagar}, {Semkov}, {Smart}, {Smith}, {Strigachev}, {Takalo}, {Tavani},
  {Tornikoski}, {Trigilio}, {Uckert}, {Umana}, {Valcheva}, {Vercellone},
  {Volvach}, \& {Wiesemeyer}}]{rai08c}
{Raiteri}, C.~M., {Villata}, M., {Larionov}, V.~M., {et~al.}
  2008{\natexlab{b}}, \aap, 491, 755

\bibitem[{{Raiteri} {et~al.}(2007){Raiteri}, {Villata}, {Larionov}, {Pursimo},
  {Ibrahimov}, {Nilsson}, {Aller}, {Kurtanidze}, {Foschini}, {Ohlert},
  {Papadakis}, {Sumitomo}, {Volvach}, {Aller}, {Arkharov}, {Bach}, {Berdyugin},
  {B{\"o}ttcher}, {Buemi}, {Calcidese}, {Charlot}, {Delgado S{\'a}nchez}, {di
  Paola}, {Djupvik}, {Dolci}, {Efimova}, {Fan}, {Forn{\'e}}, {Gomez}, {Gupta},
  {Hagen-Thorn}, {Hooks}, {Hovatta}, {Ishii}, {Kamada}, {Konstantinova},
  {Kopatskaya}, {Kovalev}, {Kovalev}, {L{\"a}hteenm{\"a}ki}, {Lanteri}, {Le
  Campion}, {Lee}, {Leto}, {Lin}, {Lindfors}, {Mingaliev}, {Mizoguchi},
  {Nicastro}, {Nikolashvili}, {Nishiyama}, {{\"O}stman}, {Ovcharov},
  {P{\"a}{\"a}kk{\"o}nen}, {Pasanen}, {Pian}, {Rector}, {Ros}, {Sadakane},
  {Selj}, {Semkov}, {Sharapov}, {Somero}, {Stanev}, {Strigachev}, {Takalo},
  {Tanaka}, {Tavani}, {Torniainen}, {Tornikoski}, {Trigilio}, {Umana},
  {Vercellone}, {Valcheva}, {Volvach}, \& {Yamanaka}}]{rai07b}
{Raiteri}, C.~M., {Villata}, M., {Larionov}, V.~M., {et~al.} 2007, \aap, 473,
  819

\bibitem[{{Raiteri} {et~al.}(1999){Raiteri}, {Villata}, {Tosti}, {Fiorucci},
  {Ghisellini}, {Takalo}, {Sillanp{\"a}{\"a}}, {Valtaoja}, {Ter{\"a}sranta},
  {Tornikoski}, {Aller}, {Aller}, {De Francesco}, {Hein{\"a}m{\"a}ki},
  {Katajainen}, {Lanteri}, {Nilsson}, {Pursimo}, {Rizzi}, \& {Sobrito}}]{rai99}
{Raiteri}, C.~M., {Villata}, M., {Tosti}, G., {et~al.} 1999, \aap, 352, 19

\bibitem[{{Raiteri} {et~al.}(2003){Raiteri}, {Villata}, {Tosti}, {Nesci},
  {Massaro}, {Aller}, {Aller}, {Ter{\"a}sranta}, {Kurtanidze}, {Nikolashvili},
  {Ibrahimov}, {Papadakis}, {Krichbaum}, {Kraus}, {Witzel}, {Ungerechts},
  {Lisenfeld}, {Bach}, {Cim{\`o}}, {Ciprini}, {Fuhrmann}, {Kimeridze},
  {Lanteri}, {Maesano}, {Montagni}, {Nucciarelli}, \& {Ostorero}}]{rai03}
{Raiteri}, C.~M., {Villata}, M., {Tosti}, G., {et~al.} 2003, \aap, 402, 151

\bibitem[{{Roming} {et~al.}(2005){Roming}, {Kennedy}, {Mason}, {Nousek}, {Ahr},
  {Bingham}, {Broos}, {Carter}, {Hancock}, {Huckle}, {Hunsberger}, {Kawakami},
  {Killough}, {Koch}, {McLelland}, {Smith}, {Smith}, {Soto}, {Boyd},
  {Breeveld}, {Holland}, {Ivanushkina}, {Pryzby}, {Still}, \& {Stock}}]{rom05}
{Roming}, P.~W.~A., {Kennedy}, T.~E., {Mason}, K.~O., {et~al.} 2005, Space
  Science Reviews, 120, 95

\bibitem[{{Sasada} {et~al.}(2010){Sasada}, {Uemura}, {Arai}, {Kukazawa},
  {Kawabata}, {Ohsugi}, {Yamashita}, {Isogai}, {Nagae}, {Uehara}, {Mizuno},
  {Katagiri}, {Takahashi}, {Sato}, \& {Kino}}]{sas10}
{Sasada}, M., {Uemura}, M., {Arai}, A., {et~al.} 2010, \pasj, 62, 645

\bibitem[{{Sikora} {et~al.}(2008){Sikora}, {Moderski}, \& {Madejski}}]{sik08}
{Sikora}, M., {Moderski}, R., \& {Madejski}, G.~M. 2008, \apj, 675, 71

\bibitem[{{Striani} {et~al.}(2010){Striani}, {Vercellone}, {Tavani},
  {Vittorini}, {D'Ammando}, {Donnarumma}, {Pacciani}, {Pucella}, {Bulgarelli},
  {Trifoglio}, {Gianotti}, {Giommi}, {Argan}, {Barbiellini}, {Caraveo},
  {Cattaneo}, {Chen}, {Costa}, {De Paris}, {Del Monte}, {Di Cocco},
  {Evangelista}, {Feroci}, {Ferrari}, {Fiorini}, {Fuschino}, {Galli},
  {Giuliani}, {Giusti}, {Labanti}, {Lazzarotto}, {Lipari}, {Longo},
  {Marisaldi}, {Mereghetti}, {Morelli}, {Moretti}, {Morselli}, {Pellizzoni},
  {Perotti}, {Piano}, {Picozza}, {Pilia}, {Prest}, {Rapisarda}, {Rappoldi},
  {Sabatini}, {Scalise}, {Soffitta}, {Trois}, {Vallazza}, {Zambra}, {Zanello},
  {Pittori}, {Verrecchia}, {Santolamazza}, {Lucarelli}, {Colafrancesco},
  {Antonelli}, \& {Salotti}}]{str10}
{Striani}, E., {Vercellone}, S., {Tavani}, M., {et~al.} 2010, \apj, 718, 455

\bibitem[{{Tavecchio} {et~al.}(2010){Tavecchio}, {Ghisellini}, {Bonnoli}, \&
  {Ghirlanda}}]{tav10}
{Tavecchio}, F., {Ghisellini}, G., {Bonnoli}, G., \& {Ghirlanda}, G. 2010,
  \mnras, 405, L94

\bibitem[{{Urry} \& {Padovani}(1995)}]{urr95}
{Urry}, C.~M. \& {Padovani}, P. 1995, \pasp, 107, 803

\bibitem[{{Vercellone} {et~al.}(2008){Vercellone}, {Chen}, {Giuliani},
  {Bulgarelli}, {Donnarumma}, {Lapshov}, {Tavani}, {Argan}, {Barbiellini},
  {Caraveo}, {Cocco}, {Costa}, {D'Ammando}, {Del Monte}, {De Paris}, {Di
  Cocco}, {Evangelista}, {Feroci}, {Fiorini}, {Froysland}, {Fuschino}, {Galli},
  {Gianotti}, {Labanti}, {Lazzarotto}, {Lipari}, {Longo}, {Marisaldi}, {Mauri},
  {Mereghetti}, {Morselli}, {Pacciani}, {Pellizzoni}, {Perotti}, {Picozza},
  {Prest}, {Pucella}, {Rapisarda}, {Soffitta}, {Trifoglio}, {Trois},
  {Vallazza}, {Vittorini}, {Zambra}, {Zanello}, {Pittori}, {Verrecchia},
  {Gasparrini}, {Cutini}, {Giommi}, {Antonelli}, {Colafrancesco}, \&
  {Salotti}}]{ver08}
{Vercellone}, S., {Chen}, A.~W., {Giuliani}, A., {et~al.} 2008, \apjl, 676, L13

\bibitem[{{Vercellone} {et~al.}(2009){Vercellone}, {Chen}, {Vittorini},
  {Giuliani}, {D'Ammando}, {Tavani}, {Donnarumma}, {Pucella}, {Raiteri},
  {Villata}, {Chen}, {Tosti}, {Impiombato}, {Romano}, {Belfiore}, {DeLuca},
  {Novara}, {Senziani}, {Bazzano}, {Fiocchi}, {Ubertini}, {Ferrari}, {Argan},
  {Barbiellini}, {Boffelli}, {Bulgarelli}, {Caraveo}, {Cattaneo}, {Cocco},
  {Costa}, {Monte}, {DeParis}, {Cocco}, {Evangelista}, {Feroci}, {Fiorini},
  {Fornari}, {Froysland}, {Fuschino}, {Galli}, {Gianotti}, {Labanti},
  {Lapshov}, {Lazzarotto}, {Lipari}, {Longo}, {Marisaldi}, {Mereghetti},
  {Morselli}, {Pellizzoni}, {Pacciani}, {Perotti}, {Picozza}, {Prest},
  {Rapisarda}, {Rappoldi}, {Soffitta}, {Trifoglio}, {Trois}, {Vallazza},
  {Zambra}, {Zanello}, {Pittori}, {Verrecchia}, {Santolamazza}, {Preger},
  {Gasparrini}, {Cutini}, {Giommi}, {Colafrancesco}, \& {Salotti}}]{ver09}
{Vercellone}, S., {Chen}, A.~W., {Vittorini}, V., {et~al.} 2009, \apj, 690,
  1018

\bibitem[{{Vercellone} {et~al.}(2010){Vercellone}, {D'Ammando}, {Vittorini},
  {Donnarumma}, {Pucella}, {Tavani}, {Ferrari}, {Raiteri}, {Villata}, {Romano},
  {Krimm}, {Tiengo}, {Chen}, {Giovannini}, {Venturi}, {Giroletti}, {Kovalev},
  {Sokolovsky}, {Pushkarev}, {Lister}, {Argan}, {Barbiellini}, {Bulgarelli},
  {Caraveo}, {Cattaneo}, {Cocco}, {Costa}, {Del Monte}, {De Paris}, {Di Cocco},
  {Evangelista}, {Feroci}, {Fiorini}, {Fornari}, {Froysland}, {Fuschino},
  {Galli}, {Gianotti}, {Labanti}, {Lapshov}, {Lazzarotto}, {Lipari}, {Longo},
  {Giuliani}, {Marisaldi}, {Mereghetti}, {Morselli}, {Pellizzoni}, {Pacciani},
  {Perotti}, {Piano}, {Picozza}, {Pilia}, {Prest}, {Rapisarda}, {Rappoldi},
  {Sabatini}, {Soffitta}, {Striani}, {Trifoglio}, {Trois}, {Vallazza},
  {Zambra}, {Zanello}, {Pittori}, {Verrecchia}, {Santolamazza}, {Giommi},
  {Colafrancesco}, {Salotti}, {Agudo}, {Aller}, {Aller}, {Arkharov}, {Bach},
  {Bachev}, {Beltrame}, {Ben{\'{\i}}tez}, {B{\"o}ttcher}, {Buemi}, {Calcidese},
  {Capezzali}, {Carosati}, {Chen}, {Da Rio}, {Di Paola}, {Dolci}, {Dultzin},
  {Forn{\'e}}, {G{\'o}mez}, {Gurwell}, {Hagen-Thorn}, {Halkola}, {Heidt},
  {Hiriart}, {Hovatta}, {Hsiao}, {Jorstad}, {Kimeridze}, {Konstantinova},
  {Kopatskaya}, {Koptelova}, {Kurtanidze}, {L{\"a}hteenm{\"a}ki}, {Larionov},
  {Leto}, {Ligustri}, {Lindfors}, {Lopez}, {Marscher}, {Mujica},
  {Nikolashvili}, {Nilsson}, {Mommert}, {Palma}, {Pasanen}, {Roca-Sogorb},
  {Ros}, {Roustazadeh}, {Sadun}, {Saino}, {Sigua}, {Sorcia}, {Takalo},
  {Tornikoski}, {Trigilio}, {Turchetti}, \& {Umana}}]{ver10}
{Vercellone}, S., {D'Ammando}, F., {Vittorini}, V., {et~al.} 2010, \apj, 712,
  405

\bibitem[{{Villata} \& {Raiteri}(1999)}]{vil99}
{Villata}, M. \& {Raiteri}, C.~M. 1999, \aap, 347, 30

\bibitem[{{Villata} {et~al.}(2007){Villata}, {Raiteri}, {Aller}, {Bach},
  {Ibrahimov}, {Kovalev}, {Kurtanidze}, {Larionov}, {Lee}, {Leto},
  {L{\"a}hteenm{\"a}ki}, {Nilsson}, {Pursimo}, {Ros}, {Sumitomo}, {Volvach},
  {Aller}, {Arai}, {Buemi}, {Coloma}, {Doroshenko}, {Efimov}, {Fuhrmann},
  {Hagen-Thorn}, {Kamada}, {Katsuura}, {Konstantinova}, {Kopatskaya}, {Kotaka},
  {Kovalev}, {Kurosaki}, {Lanteri}, {Larionova}, {Mingaliev}, {Mizoguchi},
  {Nakamura}, {Nikolashvili}, {Nishiyama}, {Sadakane}, {Sergeev}, {Sigua},
  {Sillanp{\"a}{\"a}}, {Smart}, {Takalo}, {Tanaka}, {Tornikoski}, {Trigilio},
  \& {Umana}}]{vil07}
{Villata}, M., {Raiteri}, C.~M., {Aller}, M.~F., {et~al.} 2007, \aap, 464, L5

\bibitem[{{Villata} {et~al.}(2006){Villata}, {Raiteri}, {Balonek}, {Aller},
  {Jorstad}, {Kurtanidze}, {Nicastro}, {Nilsson}, {Aller}, {Arai}, {Arkharov},
  {Bach}, {Ben{\'{\i}}tez}, {Berdyugin}, {Buemi}, {B{\"o}ttcher}, {Carosati},
  {Casas}, {Caulet}, {Chen}, {Chiang}, {Chou}, {Ciprini}, {Coloma}, {di Rico},
  {D{\'{\i}}az}, {Efimova}, {Forsyth}, {Frasca}, {Fuhrmann}, {Gadway}, {Gupta},
  {Hagen-Thorn}, {Harvey}, {Heidt}, {Hernandez-Toledo}, {Hroch}, {Hu}, {Hudec},
  {Ibrahimov}, {Imada}, {Kamata}, {Kato}, {Katsuura}, {Konstantinova},
  {Kopatskaya}, {Kotaka}, {Kovalev}, {Kovalev}, {Krichbaum}, {Kubota},
  {Kurosaki}, {Lanteri}, {Larionov}, {Larionova}, {Laurikainen}, {Lee}, {Leto},
  {L{\"a}hteenm{\"a}ki}, {L{\'o}pez-Cruz}, {Marilli}, {Marscher}, {McHardy},
  {Mondal}, {Mullan}, {Napoleone}, {Nikolashvili}, {Ohlert}, {Postnikov},
  {Pursimo}, {Ragni}, {Ros}, {Sadakane}, {Sadun}, {Savolainen}, {Sergeeva},
  {Sigua}, {Sillanp{\"a}{\"a}}, {Sixtova}, {Sumitomo}, {Takalo},
  {Ter{\"a}sranta}, {Tornikoski}, {Trigilio}, {Umana}, {Volvach}, {Voss}, \&
  {Wortel}}]{vil06}
{Villata}, M., {Raiteri}, C.~M., {Balonek}, T.~J., {et~al.} 2006, \aap, 453,
  817

\bibitem[{{Villata} {et~al.}(2009{\natexlab{a}}){Villata}, {Raiteri},
  {Gurwell}, {Larionov}, {Kurtanidze}, {Aller}, {L{\"a}hteenm{\"a}ki}, {Chen},
  {Nilsson}, {Agudo}, {Aller}, {Arkharov}, {Bach}, {Bachev}, {Beltrame},
  {Ben{\'{\i}}tez}, {Buemi}, {B{\"o}ttcher}, {Calcidese}, {Capezzali},
  {Carosati}, {da Rio}, {di Paola}, {Dolci}, {Dultzin}, {Forn{\'e}},
  {G{\'o}mez}, {Hagen-Thorn}, {Halkola}, {Heidt}, {Hiriart}, {Hovatta},
  {Hsiao}, {Jorstad}, {Kimeridze}, {Konstantinova}, {Kopatskaya}, {Koptelova},
  {Leto}, {Ligustri}, {Lindfors}, {Lopez}, {Marscher}, {Mommert}, {Mujica},
  {Nikolashvili}, {Palma}, {Pasanen}, {Roca-Sogorb}, {Ros}, {Roustazadeh},
  {Sadun}, {Saino}, {Sigua}, {Sorcia}, {Takalo}, {Tornikoski}, {Trigilio},
  {Turchetti}, \& {Umana}}]{vil09b}
{Villata}, M., {Raiteri}, C.~M., {Gurwell}, M.~A., {et~al.} 2009{\natexlab{a}},
  \aap, 504, L9

\bibitem[{{Villata} {et~al.}(2004){Villata}, {Raiteri}, {Kurtanidze},
  {Nikolashvili}, {Ibrahimov}, {Papadakis}, {Tosti}, {Hroch}, {Takalo},
  {Sillanp{\"a}{\"a}}, {Hagen-Thorn}, {Larionov}, {Schwartz}, {Basler},
  {Brown}, {Balonek}, {Ben{\'{\i}}tez}, {Ram{\'{\i}}rez}, {Sadun}, {Boltwood},
  {Carini}, {Barnaby}, {Coloma}, {Ros}, {Dai}, {Xie}, {Mattox}, {Rodriguez},
  {Asfandiyarov}, {Atkerson}, {Beem}, {Bloom}, {Chanturiya}, {Ciprini},
  {Crapanzano}, {de Diego}, {Efimova}, {Gardiol}, {Guerra}, {Kahharov},
  {Kapanadze}, {Karttunen}, {Kato}, {Kimeridze}, {Kudryavtseva}, {Lainela},
  {Lanteri}, {Larionova}, {Maesano}, {Marchili}, {Massone}, {Monroe},
  {Montagni}, {Nesci}, {Nilsson}, {Noble}, {Nucciarelli}, {Ostorero},
  {Papamastorakis}, {Pasanen}, {Peters}, {Pursimo}, {Reig}, {Ryle}, {Sclavi},
  {Sigua}, {Uemura}, \& {Wills}}]{vil04a}
{Villata}, M., {Raiteri}, C.~M., {Kurtanidze}, O.~M., {et~al.} 2004, \aap, 421,
  103

\bibitem[{{Villata} {et~al.}(2002){Villata}, {Raiteri}, {Kurtanidze},
  {Nikolashvili}, {Ibrahimov}, {Papadakis}, {Tsinganos}, {Sadakane}, {Okada},
  {Takalo}, {Sillanp{\"a}{\"a}}, {Tosti}, {Ciprini}, {Frasca}, {Marilli},
  {Robb}, {Noble}, {Jorstad}, {Hagen-Thorn}, {Larionov}, {Nesci}, {Maesano},
  {Schwartz}, {Basler}, {Gorham}, {Iwamatsu}, {Kato}, {Pullen},
  {Ben{\'{\i}}tez}, {de Diego}, {Moilanen}, {Oksanen}, {Rodriguez}, {Sadun},
  {Kelly}, {Carini}, {Miller}, {Catalano}, {Dultzin-Hacyan}, {Fan}, {Ishioka},
  {Karttunen}, {Kein{\"a}nen}, {Kudryavtseva}, {Lainela}, {Lanteri},
  {Larionova}, {Matsumoto}, {Mattox}, {Montagni}, {Nucciarelli}, {Ostorero},
  {Papamastorakis}, {Pasanen}, {Sobrito}, \& {Uemura}}]{vil02}
{Villata}, M., {Raiteri}, C.~M., {Kurtanidze}, O.~M., {et~al.} 2002, \aap, 390,
  407

\bibitem[{{Villata} {et~al.}(2008){Villata}, {Raiteri}, {Larionov},
  {Kurtanidze}, {Nilsson}, {Aller}, {Tornikoski}, {Volvach}, {Aller},
  {Arkharov}, {Bach}, {Beltrame}, {Bhatta}, {Buemi}, {B{\"o}ttcher},
  {Calcidese}, {Carosati}, {Castro-Tirado}, {da Rio}, {di Paola}, {Dolci},
  {Forn{\'e}}, {Frasca}, {Hagen-Thorn}, {Heidt}, {Hiriart}, {Jel{\'{\i}}nek},
  {Kimeridze}, {Konstantinova}, {Kopatskaya}, {Lanteri}, {Leto}, {Ligustri},
  {Lindfors}, {L{\"a}hteenm{\"a}ki}, {Marilli}, {Nieppola}, {Nikolashvili},
  {Pasanen}, {Ragozzine}, {Ros}, {Sigua}, {Smart}, {Sorcia}, {Takalo},
  {Tavani}, {Trigilio}, {Turchetti}, {Uckert}, {Umana}, {Vercellone}, \&
  {Webb}}]{vil08}
{Villata}, M., {Raiteri}, C.~M., {Larionov}, V.~M., {et~al.} 2008, \aap, 481,
  L79

\bibitem[{{Villata} {et~al.}(2009{\natexlab{b}}){Villata}, {Raiteri},
  {Larionov}, {Nikolashvili}, {Aller}, {Bach}, {Carosati}, {Hroch},
  {Ibrahimov}, {Jorstad}, {Kovalev}, {L{\"a}hteenm{\"a}ki}, {Nilsson},
  {Ter{\"a}sranta}, {Tosti}, {Aller}, {Arkharov}, {Berdyugin}, {Boltwood},
  {Buemi}, {Casas}, {Charlot}, {Coloma}, {di Paola}, {di Rico}, {Kimeridze},
  {Konstantinova}, {Kopatskaya}, {Kovalev}, {Kurtanidze}, {Lanteri},
  {Larionova}, {Larionova}, {Le Campion}, {Leto}, {Lindfors}, {Marscher},
  {Marshall}, {McFarland}, {McHardy}, {Miller}, {Nucciarelli}, {Osterman},
  {Pasanen}, {Pursimo}, {Ros}, {Sadun}, {Sigua}, {Sixtova}, {Takalo},
  {Tornikoski}, {Trigilio}, {Umana}, {Xie}, {Zhang}, \& {Zhou}}]{vil09a}
{Villata}, M., {Raiteri}, C.~M., {Larionov}, V.~M., {et~al.}
  2009{\natexlab{b}}, \aap, 501, 455

\bibitem[{{Wills} {et~al.}(1995){Wills}, {Thompson}, {Han}, {Netzer}, {Wills},
  {Baldwin}, {Ferland}, {Browne}, \& {Brotherton}}]{wil95}
{Wills}, B.~J., {Thompson}, K.~L., {Han}, M., {et~al.} 1995, \apj, 447, 139

\bibitem[{{Wilms} {et~al.}(2000){Wilms}, {Allen}, \& {McCray}}]{wil00}
{Wilms}, J., {Allen}, A., \& {McCray}, R. 2000, \apj, 542, 914

\end{thebibliography}
\end{document}